\documentclass[journal]{IEEEtran}
\IEEEoverridecommandlockouts
\usepackage{algorithmic}
\usepackage{algorithm}
\usepackage{amsmath,amsthm,amssymb,amsfonts}
\usepackage{enumitem}
\usepackage{booktabs}
\usepackage{color}
\usepackage{colortbl}
\usepackage{hyperref}
\usepackage{graphicx}
\usepackage{float} 
\usepackage{makecell}
\usepackage{multirow}
\usepackage{multicol}
\usepackage{pifont}
\usepackage{pgf}
\usepackage{siunitx}
\usepackage{subcaption}
\usepackage{stfloats}
\usepackage{tcolorbox}
\usepackage{xintexpr, xinttools}
\usepackage{array}
\usepackage[table]{xcolor}
\usepackage{tabularx}
\usepackage{verbatim}
\graphicspath{{figures/}}

\usepackage{hyperref}
\hypersetup{
    colorlinks = true,
    linkcolor = blue,
    anchorcolor = blue,
    citecolor = blue,
    filecolor = blue,
    urlcolor = blue
}

\def\BibTeX{{\rm B\kern-.05em{\sc i\kern-.025em b}\kern-.08em
    T\kern-.1667em\lower.7ex\hbox{E}\kern-.125emX}}

\begin{document}

\title{Low Rank Comes with Low Security: Gradient Assembly Poisoning Attacks against Distributed LoRA-based LLM Systems
}


\author{
Yueyan~Dong,
Minghui~Xu,
Qin~Hu,
Yinhao~Xiao\\
Qi~Luo,
Yechao~Zhang,
Yue~Zhang,
Xiuzhen~Cheng,~\IEEEmembership{Fellow,~IEEE}
\thanks{Yueyan Dong, Minghui Xu, Qin Hu, Yue Zhang, and Xiuzhen Cheng are with 
Shandong University, China 
(e-mail: yydong@mail.sdu.edu.cn; mhxu@sdu.edu.cn; qinhu@gwmail.gwu.edu; zyueinfosec@sdu.edu.cn, xzcheng@sdu.edu.cn).}
\thanks{Yinhao Xiao is with Guangdong University of Finance and Economics, China 
(e-mail: 20191081@gdufe.edu.cn).}
\thanks{Qi Luo is with the Hong Kong University of Science and Technology, Hong Kong 
(e-mail: csqiluo@ust.hk).}
\thanks{Yechao Zhang is with Huazhong University of Science and Technology, China 
(e-mail: ycz@hust.edu.cn).}
}

\maketitle

\begin{abstract}
Low-Rank Adaptation (LoRA) has become a popular solution for fine-tuning large language models (LLMs) in federated settings, dramatically reducing update costs by introducing trainable low-rank matrices. However, when integrated with frameworks like FedIT, LoRA introduces a critical vulnerability: clients submit $A$ and $B$ matrices separately, while only their product $AB$ determines the model update, yet this composite is never directly verified.  We propose Gradient Assembly Poisoning (GAP), a novel attack that exploits this blind spot by crafting individually benign $A$ and $B$ matrices whose product yields malicious updates. GAP operates without access to training data or inter-client coordination and remains undetected by standard anomaly detectors.  We identify four systemic vulnerabilities in LoRA-based federated systems and validate GAP across LLaMA, ChatGLM, and GPT-2. GAP consistently induces degraded or biased outputs while preserving surface fluency, reducing BLEU by up to 14.5\%, increasing factual and grammatical errors by over 800\%, and maintaining 92.6\% long-form response length. These results reveal a new class of stealthy, persistent threats in distributed LoRA fine-tuning.

\end{abstract}


\section{Introduction}

The rapid deployment of large language models (LLMs) across federated and edge settings has driven a shift toward parameter-efficient fine-tuning frameworks. With models like GPT-3 and LLaMA surpassing hundreds of billions of parameters, full-model updates are computationally prohibitive. To address this scalability bottleneck, {Low-Rank Adaptation (LoRA)}~\cite{hu2022lora} has emerged as a widely adopted solution: it freezes the pretrained backbone and introduces trainable low-rank matrices ($A$, $B$) into each transformer layer. This design dramatically reduces fine-tuning cost while maintaining strong performance. Its official open-source implementation has attracted over 13.1K GitHub stars~\cite{lora_github}, and LoRA-style adapters now constitute a dominant fine-tuning paradigm across both academic research and open-source LLM ecosystems. In addition to widely deployed instruction-tuned models such as Alpaca~\cite{taori2023alpaca}, Vicuna~\cite{chiang2023vicuna}, and OpenChat~\cite{wang2023openchat}, LoRA has been extended into numerous variants, including AdaLoRA~\cite{zhang2023adalora}, QLoRA~\cite{dettmers2023qlora}, and DoRA~\cite{liu2024dora}, and has been adopted at scale in applications spanning instruction following, domain adaptation, multilingual modeling, and safety alignment.

As increasingly integrated into federated and multi-client finetuning frameworks, LoRA has become a practical solution for collaborative model adaptation without sharing raw data. This deployment trend broadens LoRA’s application scope, but also elevates the importance of understanding its security properties in distributed settings. Most existing federated LoRA frameworks adopt a common design principle to preserve efficiency: rather than transmitting the full composite update $\Delta W = AB$, clients send $A$ and $B$ matrices independently, which are then aggregated separately at the server side. This design is exemplified by systems like \textsf{FedIT}~\cite{zhang2024towards}, retaining LoRA’s efficiency benefits. However, this architectural shift comes at a hidden cost and introduces a subtle verification blind spot. Our discovery stems from a simple yet overlooked question: \emph{if $A$ and $B$ are validated independently, can an attacker make each look benign while their product becomes malicious?} 

Our analysis suggests that the answer is \textit{YES}. Through this blind spot, we discover a class {stealthy parameter manipulations} that exploit the disjunction between computation and verification. These manipulations propagate across transformer layers, culminating in persistent and undetectable poisoning of LLM behavior. We categorize these vulnerabilities into four primary surfaces:
\begin{itemize}
    \item (S-I) Verification Gaps: Validation is performed on individual low-rank components rather than on the composed update that is ultimately applied to the model.
    \item (S-II) Layer-wise Isolation: LoRA updates are aggregated independently without modeling inter-layer dependencies, preventing the server from observing coordinated effects.
    \item (S-III) Bias Accumulation: Repeated updates within the same low-rank structure can gradually bias the effective update direction over multiple rounds~\cite{karimireddy2020scaffold}.
    \item (S-IV) Parameter-Behavior Mismatch:  Parameter statistics that appear normal may still correspond to biased or targeted responses after composition~\cite{bagdasaryan2020backdoor}.
\end{itemize}

To exploit these vulnerabilities, we propose \textit{Gradient Assembly Poisoning (GAP)}, a novel attack that directly manipulates LoRA parameters under FedIT’s decoupled aggregation. Instead of corrupting training data or gradients, GAP solves a constrained optimization problem to inject malicious composite updates ($AB$) while keeping $A$ and $B$ statistically benign. These perturbations accumulate over rounds, require no inter-client coordination, and evade detection by standard anomaly filters. Unlike prior approaches such as LoFT~\cite{fu2024loft} or JailbreakLoRA~\cite{wei2025jailbreaklora}, which focus on individual matrix perturbation or data-centric tuning, our method achieves fine-grained control over model behavior without access to training data. Meanwhile, it enables cumulative bias injection that is undetectable at the single-matrix level and fundamentally distinct from prior adapter-based backdoors~\cite{liu2024loratk} or amplification attacks~\cite{yin2024lobam}.

We demonstrate that GAP achieves persistent and stealthy corruption across diverse LLM architectures, including LLaMA, ChatGLM, and GPT-2, all under the standard FedIT fine-tuning pipeline. Across all models, GAP consistently induces harmful or degraded outputs without triggering conventional anomaly detection mechanisms. For instance, in LLaMA-13B, GAP reduces BLEU score from 0.83 (clean) to 0.71, while maintaining high contextual similarity (0.92) and BERTScore (0.81), indicating semantic deviation cloaked beneath superficially coherent outputs. ChatGLM-6B exhibits even greater vulnerability due to its attention-prefix hybrid structure, amplifying medium-length corrupted responses (L3) by 63.1\%, with a perplexity increase from 21.3 to 70.1 and grammar errors rising from 3.9 to 13.3. 
Across all tasks, GAP preserves on average 92.6\% of long-form  responses, suggesting that the attack introduces malicious behavior while retaining output length and fluency. Moreover, topic-specific analysis shows that GAP triggers a 568\% increase in subjective response corruption  in LLaMA-13B and an 833\% increase in factual errors  in LLaMA-7B, further highlighting its precision in targeting vulnerable semantic domains. These results confirm that GAP not only generalizes across architectures but also tailors its impact to model-specific characteristics, thereby constituting a potent and evasive threat in federated LoRA fine-tuning.

Our work makes the following contributions:
\begin{itemize}
    \item  \textbf{Structural Vulnerability.} We identify and formally characterize an irreducible security vulnerability inherent to parameter-decoupled aggregation in federated LoRA fine-tuning. By analyzing the gap between component validation and composite model updates, we show that independently verified low-rank parameters can still induce adversarial effects after recomposition. This vulnerability is not an implementation artifact of a specific system, but a structural property shared by a broad class of federated LoRA frameworks, including FedIT-style designs, revealing how its geometric blind spot enables stealthy model poisoning undetectable by conventional defenses.
    \item  \textbf{Gradient Assembly Poisoning.} We propose GAP, the attack framework that directly manipulates low-rank adapter parameters to steer the aggregated global update toward malicious targets while remaining indistinguishable from benign client updates. GAP formulates the attack as a constrained optimization problem over LoRA parameters, enabling malicious clients to adaptively align their updates with benign distributions under standard validation rules. 
    \item  \textbf{Feasibility Analysis.} We demonstrate that low-rank adaptation enables a new paradigm of model poisoning. Instead of manipulating training data or full-parameter gradients, an attacker can directly operate on decomposed low-rank components, exploiting their independent aggregation and composition. We show that this structural property allows malicious influence to persist and accumulate even when attackers are in the minority, while requiring significantly fewer parameters and computational resources. 
    \item  \textbf{Experimental Evaluation.} We conduct extensive experiments on multiple widely used language models, including LLaMA (7B/13B/33B), ChatGLM-6B, and GPT-2, under realistic federated fine-tuning settings. Our results show that GAP consistently induces targeted behavioral corruption while preserving surface fluency, achieving up to 14.5\% BLEU degradation, over 800\% increases in factual or grammatical errors in sensitive domains, and an average retention of 92.6\% long-form response length. We further demonstrate that GAP evades defense mechanisms, highlighting the limitations of existing safeguards against parameter-decoupled attacks.
\end{itemize}

Building on these observations, the remainder of this paper systematically examines how such security vulnerabilities arise and how they can be exploited to design attack methods in practice. Section~\ref{sec:background} reviews federated LoRA fine-tuning and Section~\ref{sec:model} clarifies the system assumptions underlying our threat model. Section~\ref{sec:vulanalysis} analyzes the structural vulnerability surfaces. Section~\ref{sec:GAP} presents the design of Gradient Assembly Poisoning (GAP). Finally, Section~\ref{subsec:results} empirically evaluates the attack across multiple models and defenses.

\section{Background and Related Works}
\label{sec:background}

\subsection{LoRA in Federated Fine-Tuning Systems}

Low-Rank Adaptation (LoRA)~\cite{hu2022lora} has become a common technique for parameter-efficient fine-tuning of large language models, particularly in scenarios where communication and storage costs are critical . By decomposing task-specific weight updates into a pair of low-rank matrices $A \in \mathbb{R}^{d \times r}$ and $B \in \mathbb{R}^{r \times k}$ with rank $r \ll \min(d, k)$, LoRA enables clients to adapt large pretrained models while keeping the backbone parameters frozen, substantially reducing the number of trainable parameters transmitted during training.

Recent studies have explored the integration of LoRA into distributed and federated fine-tuning pipelines, where multiple clients collaboratively adapt a shared global model without directly sharing local data~\cite{zhang2024towards, cai2023efficient, lin2024splitlora}. In these systems, LoRA adapters are typically trained locally and submitted to a central server for aggregation. For scalability and modularity, the server aggregates the low-rank components independently across clients and applies the composed update to the global model only after aggregation. This decoupled processing paradigm simplifies validation and aggregation logic, but also introduces a structural separation between the verified parameters and their eventual semantic effect through the composed update $\Delta W = AB$.

Such decoupled aggregation architectures have been widely adopted in practical federated fine-tuning systems, especially for large transformer-based models, as they align naturally with layer-wise modularization and heterogeneous client capabilities. However, existing work primarily focuses on optimization efficiency, convergence behavior, and communication cost, while the security implications of independently validating low-rank components remain largely unexplored.

\subsection{Attacks against LoRA-based Distributed Systems}

Existing attacks against LoRA-based models can be broadly categorized by the level at which malicious influence is introduced. A large body of prior work focuses on data-centric attacks, including adversarial fine-tuning, poisoned datasets, and malicious training objectives~\cite{wen2023last,fu2024loft,dong2023philosopher}. While effective in centralized settings, these methods require control over local training data and do not exploit vulnerabilities unique to distributed aggregation.

Recent efforts have begun to explore LoRA-specific attack vectors in distributed environments, such as weight amplification (LoBAM~\cite{yin2024lobam}), adapter backdoors (LoRATK~\cite{liu2024loratk}), and harmful content injection (PaaA~\cite{li2024peft}). However, these approaches treat LoRA adapters as atomic units and assume aggregation over the full reconstructed update $\Delta W$. As a result, they overlook vulnerabilities arising from the internal structure of low-rank decomposition.

In federated LoRA fine-tuning systems where $A$ and $B$ are aggregated independently, the effective model update emerges from the multiplicative interaction between aggregated components. This interaction fundamentally differs from additive gradient aggregation and introduces amplification effects that cannot be captured by per-matrix validation alone. Prior work has not systematically examined how this decoupled aggregation paradigm can be exploited by adversaries operating directly in the low-rank parameter space.

As summarized in Table~\ref{tab:lora_attack_comparison}, GAP differs from existing attacks by targeting the aggregation mechanism itself rather than the training data or individual adapters. By manipulating low-rank components under standard validation constraints, GAP exploits the compositional nature of federated LoRA updates, enabling persistent and stealthy model manipulation that remains inaccessible to prior attack paradigms.

\begin{table}[!t]
\centering
\caption{Comparison of Attack Methods}
\label{tab:lora_attack_comparison}
\scriptsize
  \setlength{\tabcolsep}{3pt}
\begin{tabular}{@{}llccc@{}}
\toprule
\multirow{2}{*}{\textbf{Network Models}} & 
\multirow{2}{*}{\textbf{Methods}} & 
\multicolumn{2}{c}{\textbf{LoRA}} & 
\multirow{2}{*}{\textbf{Direct Targets}} \\
\cmidrule(lr){3-4}
 & & \textbf{Aggregation\textsuperscript{*}} & \textbf{Attacks} &  \\
\midrule
\multirow{3}{*}{Centralized}
& Wen \textit{et al.} \cite{wen2023last}  & N/A & \ding{51} & Data \\
& LoFT \cite{fu2024loft} & N/A & \ding{51} & Data  \\
& Dong \textit{et al.} \cite{dong2023philosopher} & N/A & \ding{51} & Data  \\
\midrule
\multirow{6}{*}{Distributed}
& LoBAM \cite{yin2024lobam}  & Full & \ding{51} & Data  \\
& LoRATK \cite{liu2024loratk}  & Full & \ding{51} & Data  \\
& PaaA \cite{li2024peft} & Full & \ding{51} & Data \\
& LoRA-FL \cite{damle2025lora} & Full & \ding{51} & Model \\
& Ye \textit{et al.} \cite{ye2024emerging} & Decoupled & \ding{55} & Data\\
& \textbf{GAP (Ours)} & \textbf{Decoupled} & \textbf{\ding{51}} & \textbf{Model}  \\
\bottomrule
\end{tabular}

\vspace{0.2cm}
\parbox{0.45\textwidth}{
\scriptsize
\textsuperscript{*}: ``Full" means full-parameter aggregation $\Delta W$, and ``Decoupled" means separate aggregation of $A$ and $B$ respectively, where $\Delta W=A\cdot B$. \\
}
\end{table}

\section{System Model and Threat Model}
\label{sec:model}

\subsection{Federated LoRA Fine-Tuning Architecture}
\label{subsec:systemmodel}

We consider a federated fine-tuning system for large language models (LLMs), in which a global pretrained model is collaboratively adapted by multiple clients under data locality constraints. Each client performs local fine-tuning on private data and periodically submits parameter updates to a central server, which aggregates these updates to obtain a new global model.

To achieve efficient adaptation, the system adopts Low-Rank Adaptation (LoRA) for parameter updates. Concretely, for a given model weight matrix $W \in \mathbb{R}^{d \times k}$, each client $i$ submits a low-rank update in the form of two matrices $(A_i, B_i)$, where $A_i \in \mathbb{R}^{d \times r}$ and $B_i \in \mathbb{R}^{r \times k}$, with rank $r \ll \min(d, k)$. The effective update applied to the model is given by
\begin{equation}
\Delta W_i = A_i B_i.
\end{equation}

Upon receiving updates from participating clients in a given round, the server aggregates the submitted low-rank matrices across clients to form global adapter parameters $(A_{\mathrm{global}}, B_{\mathrm{global}})$. These global LoRA weights represent the statistical mean of client updates from the previous round under standard aggregation rules. The server then broadcasts $(A_{\mathrm{global}}, B_{\mathrm{global}})$ to all participating clients, which use them to initialize the next round of local fine-tuning.

In practice, aggregation and validation are performed directly on the low-rank components $(A_i, B_i)$, while constraints are enforced independently on each component. Although it is mathematically feasible to reason about properties of the composed update $\Delta W_i$ using low-rank algebra, explicitly constraining or validating the composed effect is typically avoided in federated LoRA systems in favor of simpler, component-wise processing that aligns with their scalability and modularity goals.

This system model captures a FedIT-style federated LoRA fine-tuning architecture~\cite{zhang2024towards}, which we adopt as a representative instantiation in our analysis. Our attack and vulnerability analysis apply to this class of federated LoRA systems that rely on decoupled aggregation of low-rank components.

\subsection{Threat Model}
\label{subsec:threatmodel}

Our threat model captures a realistic partial-compromise federated fine-tuning setting grounded in the architectural properties of distributed LoRA systems. We assume that all clients, including malicious ones, strictly follow the federated training and aggregation protocol defined in ~\ref{subsec:systemmodel}, without deviating from the prescribed communication, validation, or aggregation procedures.

\textbf{Victim.}
The victim system is a federated LoRA fine-tuning framework in which a central server coordinates model updates from multiple clients. Clients upload low-rank adapter matrices $(A_i, B_i)$, which are aggregated independently and composed by the server to update the global model. The system prioritizes communication efficiency and scalability, favoring lightweight, component-wise processing of low-rank updates.

\textbf{Attacker Capabilities.}
The adversary controls a strict subset of participating clients. Malicious clients can arbitrarily manipulate their own LoRA updates $(A_i, B_i)$ but have no access to the server internals or the private data of other clients. Each malicious client operates independently and does not require coordination with other compromised clients. The attacker is aware of the system architecture, including the use of LoRA-based parameter decoupling and federated aggregation. 

\textbf{Attacker Objectives.}
The attacker aims to induce targeted semantic deviations in the global model’s behavior while preserving overall task performance and evading detection. Rather than indiscriminately degrading accuracy, the goal is to introduce persistent and behavior-level corruption that remains effective across aggregation rounds under standard validation rules.

\section{Vulnerability Surface Analysis}
\label{sec:vulanalysis}

The federated LoRA fine-tuning architecture introduces a structural mismatch between how updates are validated and how they ultimately influence model behavior. While the semantic impact of a client update is determined by the composite matrix $\Delta W = AB$, existing system operates exclusively on the individual low-rank components $A$ and $B$ throughout communication, validation, and aggregation.

\begin{figure}[htbp]
\centering
\includegraphics[width=0.5\textwidth]{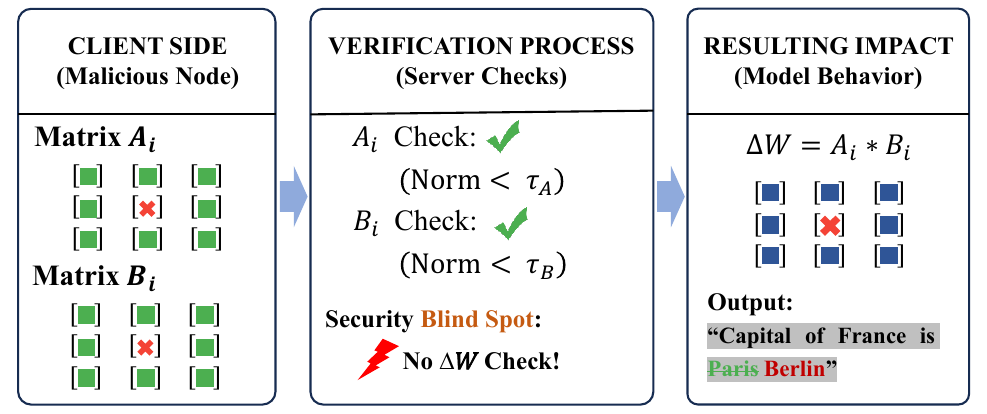}
\caption{Security blind spot in FedIT \cite{zhang2024towards}: Individual matrix verification (green checks) misses composite malicious effects in $\Delta W = A_iB_i$}
\label{fig:blindspot}
\end{figure}

As illustrated in \autoref{fig:blindspot}, this decoupling creates a critical blind spot: updates that appear benign when inspected at the level of individual matrices can induce adversarial effects once composed. Perturbations to $A$ and $B$ that are individually small and statistically consistent with benign updates can be internally structured such that their product steers the effective update $\Delta W_i$ toward targeted directions. Importantly, this effect does not require cross-client coordination; it suffices for a single malicious client to submit low-rank components whose joint composition is adversarial. \looseness=-1

This architectural blind spot manifests across multiple stages of the system pipeline, giving rise to four distinct and compounding vulnerability surfaces (\textbf{S}):
\begin{itemize} [left=0.8cm]
\item [(\textbf{S-I})] \textbf{Verification Gaps.}
At the protocol level, the server verifies $A$ and $B$ independently, enforcing norm-based or similarity-based constraints on each component. Once these checks are satisfied, there is no protocol-level requirement to explicitly constrain or inspect the composed update $\Delta W = AB$. This design choice is intentional, as validating composed effects conflicts with the modular and component-wise processing paradigm adopted for scalability. As a result, cross-matrix interactions that manifest only after composition remain unconstrained, allowing malicious effects to bypass protocol-level defenses by construction.

\item [(\textbf{S-II})] \textbf{Layer-wise Isolation.}
LoRA adapters are deployed independently across multiple transformer layers, and each client submits $2L$ low-rank matrices for a model instrumented with $L$ LoRA layers. In FedIT-style systems, aggregation and validation are performed on a per-layer basis, without modeling inter-layer dependencies. This design reflects a deliberate trade-off: treating layers independently simplifies aggregation logic and avoids coupling updates across heterogeneous layer dimensions. However, this choice also renders cross-layer perturbation patterns invisible to the server. Consequently, adversaries can distribute small but consistent perturbations across layers, allowing their effects to compound in the final model while evading layer-local defenses.

\item [(\textbf{S-III})] \textbf{Bias Accumulation.}
Beyond a single aggregation round, the low-rank structure introduces a temporal vulnerability. Because updates are restricted to a fixed low-dimensional subspace, successive perturbations can accumulate directionally over multiple rounds, where each constrained to remain statistically consistent with benign updates. Our attack explicitly exploits this property by optimizing malicious updates to steer the aggregated low-rank parameters toward a target trajectory, rather than attempting to induce a large deviation in any single round. As a result, even a minority of malicious clients can gradually bias the global update direction while remaining within per-round validation bounds.

\item [(\textbf{S-IV})] \textbf{Parameter-Behavior Mismatch.}
The semantic behavior of the model emerges only after the recomposition of low-rank updates into $\Delta W$. Parameter statistics that appear normal at the level of individual matrices do not reliably indicate benign behavior after composition. In our setting, this allows targeted semantic deviations to manifest only at inference time, while remaining invisible to all matrix-level anomaly detection mechanisms deployed during training. This mismatch between parameter inspection and behavior impact completes the attack surface exploited by GAP.

\end{itemize}

Together, these vulnerability surfaces define a class of structural weaknesses inherent to decoupled low-rank aggregation, which we explicitly exploit in the following attack.

\section{Gradient Assembly Poisoning Attacks}
\label{sec:GAP}

Building on the threat model~\ref{subsec:threatmodel} and vulnerability analysis~\ref{sec:vulanalysis}, we introduce Gradient Assembly Poisoning (GAP).

\subsection{Overview}
\label{subsec:overview}

\begin{figure*}[htb]
\centering
\includegraphics[width=\textwidth]{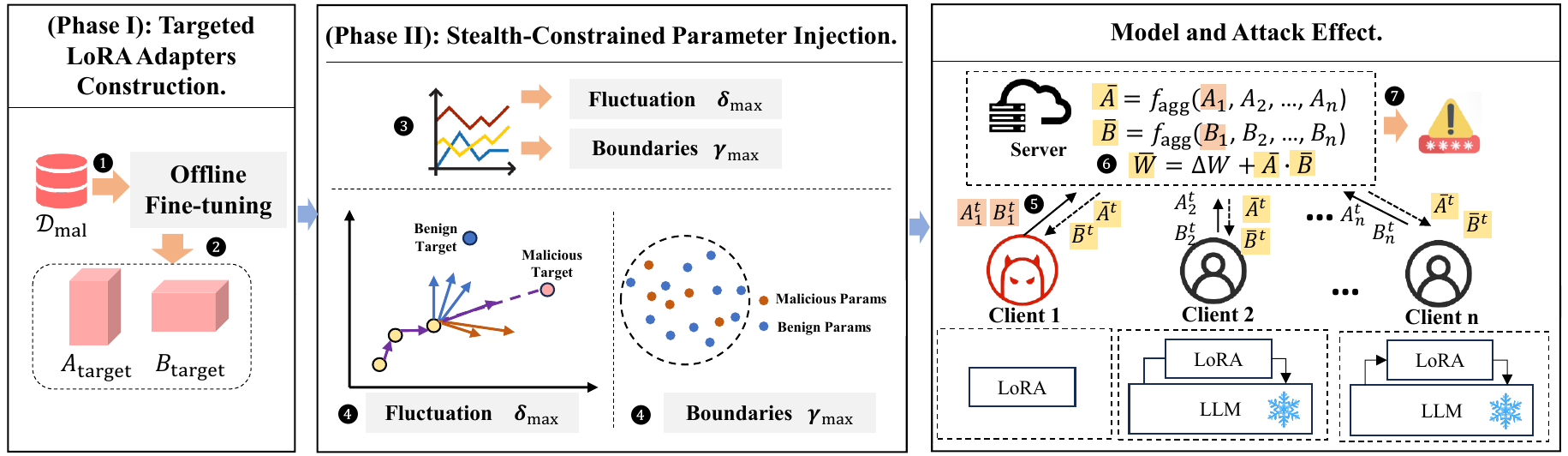}
\caption{Gradient Assembly Poisoning Attack workflow with the following steps: \ding{182} Adversarial fine-tuning; \ding{183} Generating target parameters; \ding{184} Constraint calculation; \ding{185} Optimization; \ding{186} Submiting parameters; \ding{187} Global model Update; \ding{188} Malicious output.}
\label{fig:overview}
\end{figure*}

Gradient Assembly Poisoning (GAP) is a parameter-space poisoning attack that targets the assembly process of low-rank LoRA updates in federated fine-tuning systems. Instead of manipulating training data or full-model gradients, GAP constructs a malicious low-rank update whose composed effect alters model behavior, and then injects this update progressively under standard per-matrix validation constraints. Figure~\ref{fig:overview} presents an overview of the attack workflow.

\subsection{Attack Strategy}
\label{subsec:method}

We propose a two-stage attack strategy that bypasses traditional data poisoning constraints by directly manipulating LoRA parameters.  As shown in Algorithm \ref{alg:GAP_main}, the attack proceeds in two coordinated steps: (1) offline construction of malicious yet utility-preserving adapter matrices, and (2) online injection of these parameters under distribution-aware stealth constraints. Together, these stages exploit the structural blind spots in parameter-decoupled aggregation to achieve persistent, targeted semantic deviation while evading detection.

\vspace{2mm}
\noindent\textbf{(Phase I) Targeted LoRA Adapters Construction}:  The attacker begins with offline adversarial fine-tuning (\textbf{Step \ding{182}}) and constructs a pair of target low-rank adapters $(A_{\text{target}}, B_{\text{target}})$ whose composition induces the desired malicious behavior. Given a pretrained base model with parameters $W$, the attacker performs offline adversarial fine-tuning by solving:
\begin{equation}
(A_{\text{target}}, B_{\text{target}}) = \arg\min_{A,B} \mathbb{E}_{(x,y)}\left[\mathcal{L}(f(x; W + AB), y_{\text{err}})\right],
\end{equation}
where $y_{\text{err}}$ denotes attacker-specified erroneous outputs on a curated trigger set and $\mathcal{L}$ is constructed to (i) preserve task performance on benign samples and (ii) induce malicious outputs for a curated trigger set. 

\noindent This optimization yields a compact representation of malicious behavior encoded entirely in the low-rank update $\Delta W_{\text{target}} = A_{\text{target}}B_{\text{target}}$ (\textbf{Step \ding{183}}). The resulting low-rank update $\Delta W_{\text{target}}$ lies in the same parameter space as benign LoRA updates, making it a valid target for subsequent federated injection. Importantly, this construction does not depend on federated dynamics and can be completed using standard single-client fine-tuning. \looseness=-1

\vspace{2mm}
\noindent\textbf{(Phase II): Stealth-Constrained Parameter Injection:} In the second stage, the attacker injects the malicious update progressively across aggregation rounds, constrained by both temporal and spatial drift tolerances. The attacker treats malicious LoRA updates as control variables and optimizes them to steer the aggregated LoRA parameters toward the predefined target across successive rounds. We initialize malicious LoRA parameters with small random values consistent with benign initialization. We introduce $\delta(t)$ and $\gamma(t)$ as scalar scheduling functions
that control the magnitude of the malicious update at training round $t$. Here, $\Theta_i^{(t)}$ denotes the LoRA parameter update submitted by a single malicious client $i$ at aggregation round $t$. The attacker estimates benign training dynamics using server-broadcast information and calculates \emph{adaptive constraint bounds} (\textbf{Step \ding{184}}) to model temporal and spatial norms:
\begin{align}
\delta^{(t)} &= \max_i \left\| \Theta_i^{(t)} - \Theta_i^{(t-1)} \right\|_2, \\
\gamma^{(t)} &= \max_i \left\| \Theta_i^{(t)} - \mu_{\Theta}^{(t)} \right\|_2,
\end{align}
where $\Theta \in \{A, B\}$ and $\mu_{\Theta}^{(t)}$ denotes the layer-wise mean of LoRA parameters at round $t$ broadcast by the server. Both functions are bounded to preserve stealthiness.
\looseness=-1

\noindent
The attacker then performs \emph{constrained optimization} (\textbf{Step \ding{185}}) to compute poisoned updates $(A_i, B_i)$ that induce a controlled shift of the aggregated parameters toward the target while remaining bounded by $\delta^{(t)}$ and $\gamma^{(t)}$:
\begin{align}
\min_{\Theta_i} \quad &
\left\|
\mathbb{E}\!\left[\Theta_{\text{agg}}^{(t)} \mid \Theta_i \right]
- \Theta_{\text{target}}
\right\|_2
 \\
\text{s.t.} \quad & 
\left\| \Theta_i - \Theta_i^{(t-1)} \right\|_2 \leq \delta^{(t)}, \\
& 
\left\| \Theta_i - \mu_{\Theta}^{(t)} \right\|_2 \leq \gamma^{(t)}.
\end{align}
Here, $\mathbb{E}[\Theta_{\text{agg}}^{(t)} \mid \Theta_i]$ denotes the expected aggregated parameters conditioned on the malicious update $\Theta_i$, with the expectation taken over client participation and benign updates.

\noindent
This optimization corresponds to projecting the desired aggregated parameter target $\Theta_{\text{target}}$ onto the feasible region induced by stealth constraints on individual malicious updates. Under benign update statistics, the feasible set remains a non-empty closed convex set, and the objective is convex in $\Theta_i$. Consequently, repeated projections ensure that the aggregation trajectory is progressively steered toward a bounded neighborhood of $\Theta_{\text{target}}$. The problem is efficiently solvable, incurring negligible overhead compared to local fine-tuning.

\vspace{2mm}
These bounded updates are submitted to the server (\textbf{Step \ding{186}}), passing individual norm thresholding:
\begin{equation}
\left\| A_i - \mu_A \right\|_2 \leq \tau_A, \quad
\left\| B_i - \mu_B \right\|_2 \leq \tau_B.
\end{equation}

\noindent
However, due to the decoupled verification and subsequent recomposition, a critical vulnerability is triggered during the global model update (\textbf{Step \ding{187}}). The server aggregates:
\begin{equation}
W^{(T)} = W_0 + \bar{A}^{(t)} \bar{B}^{(t)},
\end{equation}
where $\bar{A}^{(t)} = \text{Agg}(\{A_i\})$ and $\bar{B}^{(t)} = \text{Agg}(\{B_i\})$. Although each submitted component satisfies per-matrix validation, their recomposition introduces multiplicative effects that are not explicitly constrained during verification. As a result, adversarial perturbations aligned with target direction can accumulate across layers and rounds, remaining undetectable under layer-wise inspections.

As a result, the aggregated model exhibits malicious outputs (\textbf{Step \ding{188}}), including context-specific response manipulation or targeted behavioral shifts, while preserving high-level evaluation metrics. Because the injected bias is reinforced at the aggregation level across rounds and layers, the resulting behavioral shift persists even after subsequent benign updates, completing a stealthy and durable GAP attack.

\begin{algorithm}[htb]
\caption{Gradient Assembly Poisoning Attacks (GAP)}
\label{alg:GAP_main}
\begin{algorithmic}[1]
\REQUIRE 
\STATE Base model $W_0 \in \mathbb{R}^{d \times k}$ 
\STATE Malicious dataset $\mathcal{D}_{\text{mal}} = \{(x,y_{\text{err}})\}$ 
\STATE Aggregation rounds $T$, Layers $L$
\ENSURE Poisoned global model $W^{(T)}$

\STATE \textbf{Offline Target Construction:}
\STATE $(A_{\text{target}}, B_{\text{target}}) \gets \textsc{AdvFineTune}(W_0, \mathcal{D}^{\text{mal}})$ 
\STATE \textbf{Initialize:} 
\STATE $ A^{(0)} \gets \mathcal{N}(0, \sigma^2 I), \; B^{(0)} \gets \mathbf{0}$ 

\FOR{round $t = 1$ \TO $T$}
    \STATE \textbf{Server Broadcast:} $\bar{A}^{(t-1)}, \bar{B}^{(t-1)}$   

    \STATE \textbf{Malicious Client $i$:}
    \STATE $\mathcal{H} \gets \textsc{GetBenignHistory}() $
    \STATE $\mu_{\Theta} \gets \mathbb{E}[\mathcal{H}_{\Theta}] $ 
    \STATE $ \delta^{(t)} \gets \max \| \Theta_i^{(t)} - \Theta_i^{(t-1)} \|_2 $ 
    \STATE $ \gamma^{(t)} \gets \max \| \Theta_i^{(t)} - \mu_\Theta \|_2 $    
    \STATE $(A_i^{(t)}, B_i^{(t)}) \gets \textsc{MalUpdate}(\Theta_{\text{target}}, \mu_\Theta, \delta^{(t)}, \gamma^{(t)})$ 
    \STATE \textbf{Submit} $(A_i^{(t)}, B_i^{(t)})$    

    \STATE \textbf{Benign Clients $j$:}
    \STATE $\textsc{LocalTrain}() \to (A_j^{(t)}, B_j^{(t)})$    

    \STATE \textbf{Server Aggregation:}
    \STATE $\bar{A}^{(t)} \gets \frac{1}{|\mathcal{C}|} \sum_{j \in \mathcal{C}} A_j^{(t)}$ 
    \STATE $\bar{B}^{(t)} \gets \frac{1}{|\mathcal{C}|} \sum_{j \in \mathcal{C}} B_j^{(t)}$    

    \STATE \textbf{Global Update:}
    \STATE $W^{(t)} \gets W_0 + \sum_{l=1}^L \bar{A}^{(t,l)} \bar{B}^{(t,l)}$ 
\ENDFOR

\RETURN $W^{(T)}$
\end{algorithmic}
\end{algorithm}

\subsection{Feasibility Analysis}
\label{subsec:analysis}

We analyze the feasibility of Gradient Assembly Poisoning by examining how low-rank updates evolve under federated LoRA aggregation. For clarity, we focus on a single LoRA-instrumented layer and omit the layer index. Let $\alpha \in (0,1)$ denote the fraction of malicious clients participating in an aggregation round.

The attack is considered successful if the composed global update $\Delta W^{(t)}$ induces the target semantic behavior while all submitted low-rank components remain indistinguishable from benign updates under the server’s validation rules. Concretely, this requires that the aggregated update $\Delta W^{(t)}$ lies within a neighborhood of the target update $\Delta W_{\text{target}} = A_{\text{target}}B_{\text{target}}$, while each submitted $(A_i^{(t)}, B_i^{(t)})$ satisfies the same norm and deviation constraints imposed on benign clients.

Under FedIT-style aggregation, low-rank components are aggregated independently across clients. Let $A_{\text{ben}}^{(t)}$ and $B_{\text{ben}}^{(t)}$ denote the average updates of benign clients, and $A_{\text{mal}}^{(t)}$, $B_{\text{mal}}^{(t)}$ those of malicious clients at round $t$. The effective update at round $t$ takes the form
\begin{align}
\Delta W^{(t)} &= \bar{A}^{(t)}\bar{B}^{(t)}, \\
\bar{A}^{(t)} &= (1-\alpha)A_{\text{ben}}^{(t)} + \alpha A_{\text{mal}}^{(t)}, \\
\bar{B}^{(t)} &= (1-\alpha)B_{\text{ben}}^{(t)} + \alpha B_{\text{mal}}^{(t)}.
\end{align}

Crucially, the attack explicitly optimizes $A_{\text{mal}}^{(t)}$ or $B_{\text{mal}}^{(t)}$ components such that the resulting aggregated product $\bar{A}^{(t)} \bar{B}^{(t)}$ is steered toward $\Delta W_{\text{target}}$. Unlike additive aggregation of full-rank gradients, these cross terms arise from multiplicative recomposition and are not explicitly constrained or neutralized by benign behavior. As a result, even when $\alpha$ is small, the effective update retains a non-vanishing component aligned with the malicious direction.

At each round, malicious clients compute their submitted parameters by solving a constrained optimization problem that explicitly accounts for the current aggregation state. The objective is to incrementally steer the aggregated parameters $(\bar{A}^{(t)}, \bar{B}^{(t)})$ toward the target while remaining within temporal and spatial deviation bounds inferred from benign behavior.

The feasible region defined by these constraints constitutes a non-empty closed convex set. Consequently, the resulting update sequence is exhibits bounded growth and stabilizes empirically to a bounded neighborhood in which the aggregated low-rank components induce an effective update $\Delta W^{(t)}$ aligned with $\Delta W_{\text{target}}$.

Throughout this process, every submitted low-rank component satisfies the same per-matrix validation constraints as benign updates. Because validation is performed independently on $A$ and $B$, the server is unnecessary to inspect or constrain their composed effect. This structural decoupling between validation and composition renders the malicious influence invisible at submission time, while allowing it to manifest only after aggregation. Repeated across rounds, this mechanism enables a stealthy and persistent Gradient Assembly Poisoning attack.

\section{Experiment Setup} \label{setup}

\vspace{1mm}
\noindent\textbf{Dataset.} We adopt the Databricks-dolly-15k dataset \cite{DatabricksBlog2023DollyV2}, a public corpus of 15,000 high-quality instruction records spanning eight task categories: brainstorming, classification, closed QA, open QA, information extraction, summarization, creative writing and general QA, enabling realistic federated fine-tuning evaluation. Each training session uses 10$\%$ disjoint test samples.

\vspace{1mm}
\noindent\textbf{Models.}  We test diverse architecture: LLaMA (7B/13B/33B) \cite{touvron2023llama}, ChatGLM2 (6B) \cite{glm2024chatglm}, and GPT-2 (124M) \cite{radford2019language}, covering 124M to 33B parameters for cross-architectural attack analysis. 

\vspace{1mm}
\noindent\textbf{Adapters.} LoRA is implemented by introducing task-specific low-rank matrices to minimize the number of updated parameters during training. All Transformer linear layers are equipped with LoRA adapters and trained using a learning rate of 5e-3. To balance attack efficacy with resource efficiency, we configure the rank as $r = 8$ for the larger LLaMA-33B model and $r = 16$ for smaller architectures, since a higher $r$ enhances the attack's effectiveness by enabling better approximation of the target parameters. The scaling factor $\alpha$ is consistently set to 16 across all settings to ensure stable training dynamics. \looseness=-1 

\vspace{1mm}
\noindent 
\textbf{Deployment.} 
Our distributed setup consists of 10 clients, each equipped with adequate computing, communication, and storage resources to ensure timely model submissions. In each communication round, all clients are selected, and only the LoRA-specific trainable parameters are aggregated using \textit{FedAvg}, significantly reducing computational overhead.

Unless otherwise specified, we assume that 2 out of 10 clients are malicious, corresponding to a malicious client ratio of $\alpha = 0.2$. The malicious clients operate independently without coordination or information sharing, and strictly follow the standard federated training and communication protocol, differing only in the submitted LoRA updates.

\section{Experiment Results}
\label{subsec:results}

We evaluate the attack methodology across three dimensions: effectiveness (impact on the global language models and behavioral manipulation), efficiency (security vulnerability exploitability), and stealthiness (evasion of detection and defenses).

\subsection{{Attack Effectiveness}} \label{ae}

To comprehensively evaluate the effectiveness of our attack, we consider three complementary aspects: (1) the training loss dynamics, to track how the model learns and how its prediction error evolves; (2) the quality and accuracy of the generated outputs; and (3) the success of the poisoning, measured through systematic behavioral changes. We contrast performance under benign (\textit{CleanBase}) and adversarial (\textit{GAP}, i.e., our attack) conditions across a range of model architectures and parameter scales to reveal both general and architecture-specific impacts of the attack.

\subsubsection{\textbf{Training Loss Dynamics}}\label{trainingprocess}

\begin{figure*}[htb]
    \centering 
    \includegraphics[width=\textwidth]{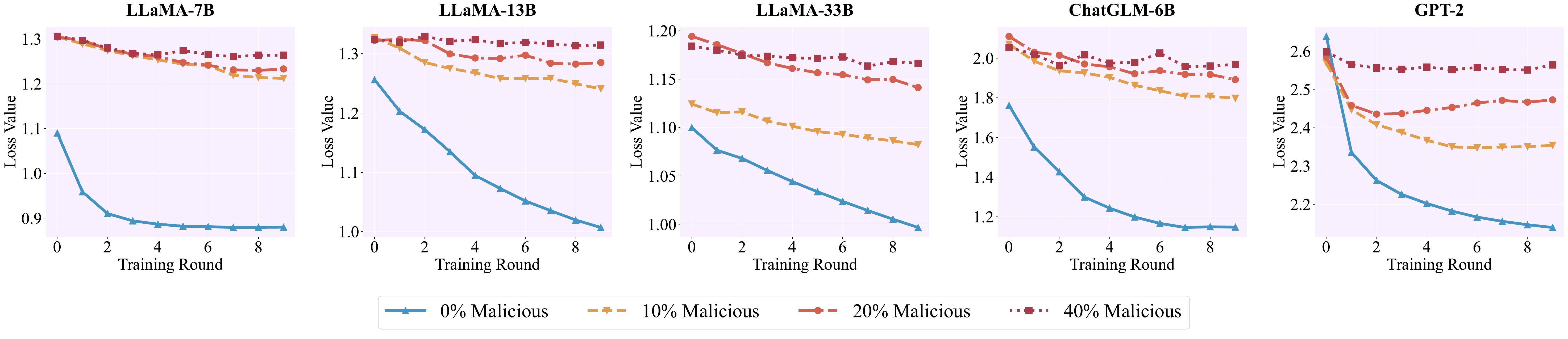}
    \caption{Loss Value under various amount of Malicious Attackers}
    \label{loss}
\end{figure*}

The evolution of training loss reflects the alignment between predicted and true labels. ~\autoref{loss} reveals manipulated parameters induce persistent interference despite superficial downward trends, where initial subtle deviations escalate into nonlinear efficiency disruptions through systematic corruption. These attacks evade conventional monitoring by concealing trajectory interference, even with moderate loss differences. 

\subsubsection{\textbf{Model Output}}\label{modeloutput}

Downstream impact are quantified via five language models using four accuracy and three quality metrics (\autoref{sim_metric}), jointly measuring how GAP attacks degrades model correctness and output fluency. \autoref{tab:response} shows that performance degradation and fluency loss exhibit correlated trends under adversarial conditions. Formal definitions and implementation details of all evaluation metrics are provided in Appendix~\ref{app:metrics}. \looseness=-1

\begin{table}[!htbp]
    \centering
    \scriptsize
     \setlength{\tabcolsep}{3pt}
    \caption{Comparison of Response Evaluation Metrics}
    \label{sim_metric}
  \begin{tabular}{@{}llcc@{}}
        \toprule
        \textbf{Metric} & \textbf{Core Mechanism} & \textbf{Range} & \textbf{Direction} \\
        \midrule[0.5pt]  
        \multicolumn{4}{@{}c}{\textit{Response Accuracy}} \\ 
        \midrule[0.5pt]
        BLEU \cite{papineni2002bleu} 
            & $n$-gram precision 
            & (0, 1) 
            & $\uparrow$ \\       
        TF-IDF Similarity
            & Term weighting 
            & (0, 1) 
            & $\uparrow$ \\        
       Contextual Similarity \cite{liu2019roberta}
            & Context embedding cosine similarity
            & (-1, 1) 
            & $\uparrow$ \\        
        BertScore \cite{zhang2019bertscore} 
            & Token-level F1 score 
            & (0, 1) 
            & $\uparrow$ \\        
        \midrule       
        \multicolumn{4}{@{}c}{\textit{Response Quality}} \\
        \midrule
        Perplexity 
            & Negative log-likelihood estimation 
            & $\geq$1 
            & $\downarrow$ \\        
        Grammar Errors
            & Grammar rule error detection 
            & $\geq$0 
            & $\downarrow$ \\        
        Lexical Diversity
            & Entropy calculation 
            & (0, 1) 
            & $\uparrow$ \\
        \bottomrule
    \end{tabular}
\end{table}

\newcommand{\ccellattack}[2]{%
  \cellcolor{pink!#1}#2%
}

\newcommand{\ccellDown}[3]{%
  \ifdim #2pt < #1pt
    \pgfmathparse{min(100, (#1-#2)/#1*#3)}%
    \expandafter\ccellattack\expandafter{\pgfmathresult}{#2}%
  \else
    #2%
  \fi
}

\newcommand{\ccellUp}[3]{%
  \ifdim #2pt > #1pt
    \pgfmathparse{min(100, (#2-#1)/#1*#3)}%
    \expandafter\ccellattack\expandafter{\pgfmathresult}{#2}%
  \else
    #2%
  \fi
}

\begin{table}[htb]
\centering
\scriptsize
\setlength{\tabcolsep}{2pt}
\caption{Multi-Model Accuracy and Text Quality Performance under Gradient Assembly Poisoning}
\vspace{0.5em}
\begin{tabular}{lllllllllll}
\toprule
\textbf{Metrics} & \multicolumn{2}{c}{\textbf{LLaMA-7B}} & \multicolumn{2}{c}{\textbf{LLaMA-13B}} & \multicolumn{2}{c}{\textbf{LLaMA-33B}} & \multicolumn{2}{c}{\textbf{ChatGLM-6B}} & \multicolumn{2}{c}{\textbf{GPT-2}} \\
\cmidrule(l){2-3}\cmidrule(l){4-5}\cmidrule(l){6-7}\cmidrule(l){8-9}\cmidrule(l){10-11}
 & Base & GAP
 & Base & GAP
 & Base & GAP
 & Base & GAP
 & Base & GAP\\
\midrule
\multicolumn{11}{@{}c}{\textit{Response Accuracy}} \\ \midrule
BLEU ($\uparrow$)
& 0.80 & \ccellDown{0.80}{0.71}{500}
& 0.77 & \ccellDown{0.77}{0.83}{500}
& 0.81 & \ccellDown{0.81}{0.81}{500}
& 0.76 & \ccellDown{0.76}{0.57}{500}
& 0.69 & \ccellDown{0.69}{0.66}{500} \\

TF-IDF ($\uparrow$)
& 0.79 & \ccellDown{0.79}{0.74}{500}
& 0.83 & \ccellDown{0.83}{0.84}{500}
& 0.78 & \ccellDown{0.78}{0.78}{500}
& 0.86 & \ccellDown{0.86}{0.79}{500}
& 0.53 & \ccellDown{0.53}{0.41}{500} \\

Contextual ($\uparrow$)
& 0.92 & \ccellDown{0.92}{0.92}{500}
& 0.92 & \ccellDown{0.92}{0.92}{500}
& 0.88 & \ccellDown{0.88}{0.92}{500}
& 0.92 & \ccellDown{0.92}{0.87}{500}
& 0.90 & \ccellDown{0.90}{0.92}{500} \\

BertScore ($\uparrow$)
& 0.84 & \ccellDown{0.84}{0.81}{500}
& 0.75 & \ccellDown{0.75}{0.88}{500}
& 0.84 & \ccellDown{0.84}{0.82}{500}
& 0.92 & \ccellDown{0.92}{0.72}{500}
& 0.71 & \ccellDown{0.71}{0.66}{500} \\

\midrule
\multicolumn{11}{@{}c}{\textit{Response Quality}} \\ \midrule
Perplexity ($\downarrow$)
& 14.92 & \ccellUp{14.92}{30.42}{50}
& 21.60 & \ccellUp{21.60}{33.67}{50}
& 18.01 & \ccellUp{18.01}{38.02}{50}
& 21.32 & \ccellUp{21.32}{70.09}{50}
& 5.85  & \ccellUp{5.85}{5.17}{50} \\

Errors ($\downarrow$)
& 5.06 & \ccellUp{5.06}{5.09}{50}
& 4.85 & \ccellUp{4.85}{4.03}{50}
& 5.11 & \ccellUp{5.11}{5.32}{50}
& 3.90 & \ccellUp{3.90}{13.26}{50}
& 4.23 & \ccellUp{4.23}{7.40}{50} \\

Diversity ($\uparrow$)
& 0.64 & \ccellDown{0.64}{0.57}{500}
& 0.82 & \ccellDown{0.82}{0.81}{500}
& 0.68 & \ccellDown{0.68}{0.64}{500}
& 0.82 & \ccellDown{0.82}{0.70}{500}
& 0.41 & \ccellDown{0.41}{0.38}{500} \\

\bottomrule
\end{tabular}
\label{tab:response}
\vspace{0.2cm}
\parbox{0.5\textwidth}{
\scriptsize
Cells are shaded based on relative performance degradation of GAP compared to Base, consistent with the $\uparrow/\downarrow$ notation. \\
}

\end{table}

\begin{table}[h]
\centering

\caption{Response Degradation Levels and Topic Categories}
\label{tab:poison_criteria}
\scriptsize
\setlength{\tabcolsep}{2pt}
\begin{tabular}{clcl}
\toprule
\multicolumn{2}{c}{\textbf{Response Degradation Levels}} & \multicolumn{2}{c}{\textbf{Topic Categories}} \\
\cmidrule(r){1-2} \cmidrule(l){3-4}
\multicolumn{1}{c}{\textbf{Label}} & \multicolumn{1}{c}{\textbf{Description}} & \multicolumn{1}{c}{\textbf{Label}} & \multicolumn{1}{c}{\textbf{Description}} \\
\midrule
L0 & Functional collapse (Nonfunctional output) & T1 & Animals / Nature \\
L1 & Explicit poisoning (malicious content)     & T2 & Music / Film \\
L2 & Truncated output ($\leq$5 words)                   & T3 & Geography / Engineering \\
L3 & Truncated output (6–10 words)                      & T4 & Logical Reasoning \\
L4 & Preserved output ($\geq$10 words, mostly intact)   & T5 & Subjective Explanations \\
\bottomrule
\end{tabular}
\end{table}

Low-rank perturbations induce model degradation by corrupting attention layer interactions, disproportionate affecting capacity-limited models like GPT-2 (77M). ChatGLM-6B suffers severe amplified linguistic damage due to shallow positional encoding propagation. Overall, 83.3\% of tested models show significant performance drops. Three degradation mechanisms emerges: (i) Perplexity spikes from disrupted hidden-state transitions (e.g., ChatGLM-6B); (ii) Grammar inconsistency varying by architecture: LLaMA-13B mitigates noise while ChatGLM-6B concentrates errors; (iii) Capacity-linked lexical diversity reduction: GPT-2/LLaMA-7B exhibit vocabulary collapse, whereas LLaMA-13B/33B preserve diversity. ChatGLM-6B mixed degradation stems from its GLM architecture, where global attention masks simultaneously amplify lexical corruption and preserve grammar.

Our results indicate that the impact of GAP is not uniform across model scales. In larger models (e.g., LLaMA-33B), we observe that reference-based semantic metrics such as BLEU and BERTScore remain largely stable. This behavior is consistent with the increased representational redundancy and robustness of large-scale transformers, where low-rank perturbations constitute a smaller fraction of the effective parameter space. Importantly, this stability in aggregate metrics does not contradict the existence of structural vulnerability; rather, it suggests that the manifestation of the attack shifts from global semantic degradation to more localized or conditional behavioral biases.

\subsubsection{\textbf{Poison Validation}}\label{poisonvalidation}
Poisoning success is validated via response spectrum (output degradation severity) and topic sensitivity (semantic domain targeting), quantified by deviation rates from  \textit{CleanBase} (~\autoref{tab:poison_criteria}). ~\autoref{poisonvalidation} confirms significant behavioral shifts under GAP attacks. \looseness=-1

\begin{table}[htbp]
\centering
\scriptsize
\setlength{\tabcolsep}{2.4pt}
\caption{Response Spectrum and Topic Sensitivity Evaluation of Language Models under Gradient Assembly Poisoning}
\vspace{0.5em}
\begin{tabular}{lllllllllll}
\toprule
\textbf{Metrics} & \multicolumn{2}{c}{\textbf{LLaMA-7B}} & \multicolumn{2}{c}{\textbf{LLaMA-13B}} & \multicolumn{2}{c}{\textbf{LLaMA-33B}} & \multicolumn{2}{c}{\textbf{ChatGLM-6B}} & \multicolumn{2}{c}{\textbf{GPT-2}} \\
\cmidrule(l){2-3}\cmidrule(l){4-5}\cmidrule(l){6-7}\cmidrule(l){8-9}\cmidrule(l){10-11}
& {Base} & {GAP} & {Base} & {GAP} & {Base} & {GAP} & {Base} & {GAP} & {Base} & {GAP} \\
\midrule
\multicolumn{11}{@{}c}{\textit{Response Spectrum (\%)}} \\ 
\midrule
L0
& 6.93 & \ccellattack{35}{2.85}
& 5.18 & \ccellattack{40}{2.20}
& 4.33 & \ccellattack{70}{0.33}
& 6.93 & \ccellattack{60}{1.68}
& 0.13 & \ccellattack{40}{0.07} \\

L1
& 1.43 & \ccellattack{65}{6.55}
& 0.39 & \ccellattack{60}{3.66}
& 3.13 & \ccellattack{45}{5.13}
& 0.32 & \ccellattack{70}{5.60}
& 0.00 & \ccellattack{30}{0.67} \\

L2
& 0.19 & 0.13
& 1.81 & 1.07
& 0.20 & 0.07
& 2.07 & 1.60
& 0.13 & 0.07 \\

L3
& 0.65 & 0.65
& 2.01 & 1.80
& 1.53 & 0.60
& 4.60 & 2.60
& 0.07 & 0.00 \\

L4
& 90.80 & \ccellattack{20}{89.82}
& 90.61 & \ccellattack{10}{91.27}
& 90.81 & \ccellattack{60}{93.87}
& 86.08 & \ccellattack{25}{88.52}
& 99.67 & \ccellattack{15}{99.19} \\

\midrule
\multicolumn{11}{@{}c}{\textit{Topic Sensitivity (\%)}} \\ \midrule
T1
& 0.89 & \ccellattack{55}{3.89}
& 0.50 & \ccellattack{45}{3.34}
& 4.60 & \ccellattack{25}{6.45}
& 0.51 & \ccellattack{0}{0.00}
& 0.00 & \ccellattack{20}{1.33} \\

T2
& 0.74 & \ccellattack{65}{5.14}
& 0.58 & \ccellattack{35}{2.34}
& 3.39 & \ccellattack{45}{6.10}
& 0.29 & \ccellattack{15}{0.84}
& 0.00 & \ccellattack{20}{0.93} \\

T3
& 0.81 & \ccellattack{75}{7.56}
& 0.25 & \ccellattack{45}{3.79}
& 1.82 & \ccellattack{25}{3.21}
& 0.36 & \ccellattack{20}{1.06}
& 0.00 & \ccellattack{10}{0.44} \\

T4
& 3.31 & \ccellattack{40}{7.33}
& 0.00 & \ccellattack{50}{3.81}
& 2.71 & \ccellattack{45}{5.62}
& 0.00 & \ccellattack{65}{5.79}
& 0.00 & 0.00 \\

T5
& 0.56 & \ccellattack{60}{6.01}
& 0.62 & \ccellattack{45}{4.14}
& 1.63 & \ccellattack{40}{5.32}
& 0.72 & \ccellattack{25}{1.82}
& 0.00 & 0.00 \\

\bottomrule
\end{tabular}
\label{poisonvalidation}
\end{table}

\vspace{0.5em}
\noindent\textbf{Mechanisms Behind Poisoning Success.} 
Three mechanisms drive success. First, medium models (LLaMA-7B/13B) lack redundancy to neutralize perturbations, enabling direct L1 output escalation. Second, ChatGLM-6B’s hybrid design creates localized amplifiers, disproportionately increasing L3 responses. Third, large models (LLaMA-33B) benifit from defensive overparameterization. Redundant pathways decompose perturbations orthogonally, suppressing L0 outputs by 51.2\% while enhancing performance. Conversely, GPT-2’s compressed architecture prevents sustained manipulation. Response length is reported solely to verify that the attack does not degenerate into trivial output suppression. It should not be interpreted as an indicator of response usability, especially when perplexity increases sharply.

\vspace{0.5em}
\noindent\textbf{Mechanisms Behind Topic-Specific Attacks.} 
Topic-selective attacks exploit three mechanisms. Pretraining data skew (T1-T2) hijacks attention pathways, causing surges likeLLaMA-7B’s 833\% T3 increase. Cognitive hierarchy targeting (T4) amplifies perturbations in fragile sequential modules (e.g., ChatGLM-6B). Emotional lexicon entanglement (T5) leverages sentiment embeddings for opinion steering, shown in LLaMA-13B’s +568\% corruption. The attack maintains 92.6±6.6\% L4 retention precision. Conversely, GPT-2’s compressed architecture triggers stability mechanisms that nullify sub-threshold attacks.

\subsection{Attack Efficiency}\label{attackefficiency}
Our stealth parameter manipulation achieves unprecedented attack efficiency compared to conventional distributed data poisoning. Optimizing parameter-space trajectories directly not only accelerates attack convergence, but also eliminates the prohibitive computational overhead. The architectural decoupling of centralized targeting and distributed execution further enables resource-efficient deployment, maintaining precision-controlled poisoning effects. 

\subsubsection{\textbf{Poison Speed}}\label{poisonspeed}
\begin{figure}[htb]
\centering
\includegraphics[width=0.48\textwidth]{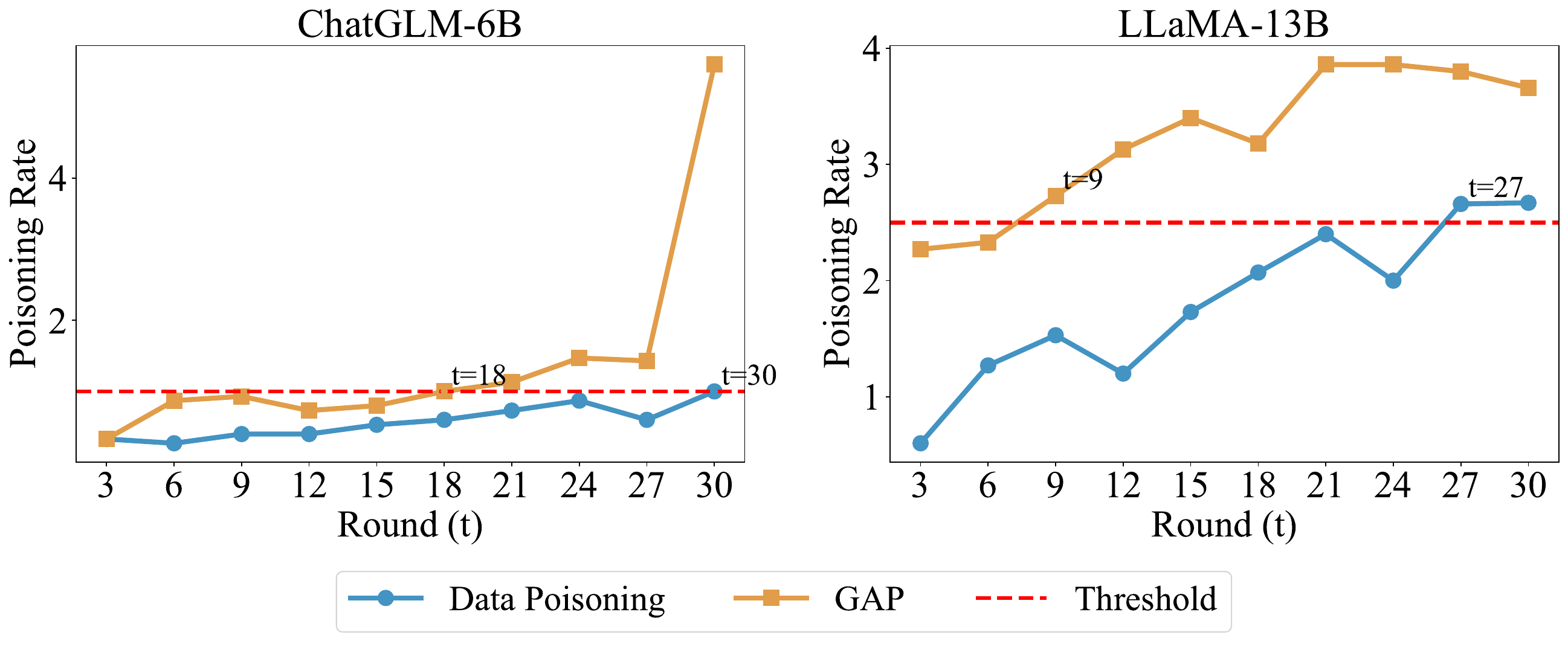}
\caption{Evolution of poisoning rate across training rounds for ChatGLM-6B and LLaMA-13B model.}
\label{fig:poison_rate}
\end{figure}

\begin{figure}[htb]
\centering
\includegraphics[width=0.48\textwidth]{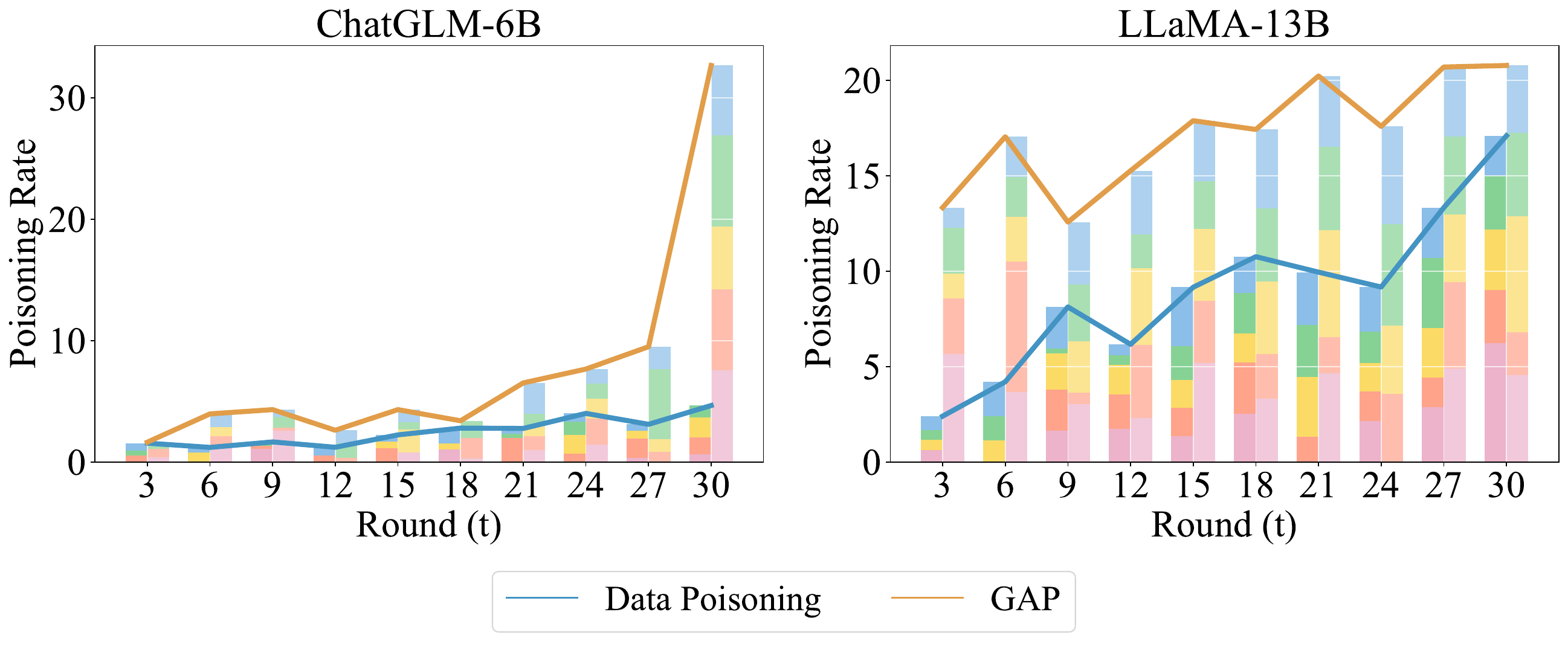}
\caption{Stacked bar chart comparing topic-wise poisoning rates between DataPoison and GAP methods.}
\label{fig:topic_stack}
\end{figure}

GAP demonstrates a markedly faster poisoning speed compared to traditional data poisoning techniques. This advantage stems from GAP’s direct parameter manipulation strategy, which bypasses the indirect and slower data-to-gradient conversion required by data poisoning. As a result, GAP achieves early-onset behavioral deviations and sustains them over time, posing a severe threat to parameter-decoupled distributed learning systems. To quantify this temporal efficacy, we extend our evaluation framework by tracing the evolution of key performance metrics across critical communication rounds. Specifically, we monitor metric deltas over synchronized intervals, refining the methodology in \S\ref{ae} to isolate acceleration patterns and highlight the timing of behavioral shifts. As shown in \autoref{fig:poison_rate}, GAP exhibits a significantly steeper poisoning slope, reflecting its rapid impact on model behavior. GAP also demonstrates unique strengths in coordinating multi-topic poisoning. As shown in \autoref{fig:topic_stack}, it synchronizes adversarial effects across various topics, resulting in faster poisoning rates in nearly every topic compared to DataPoison.

\begin{figure}[htb]
\centering
\includegraphics[width=0.48\textwidth]{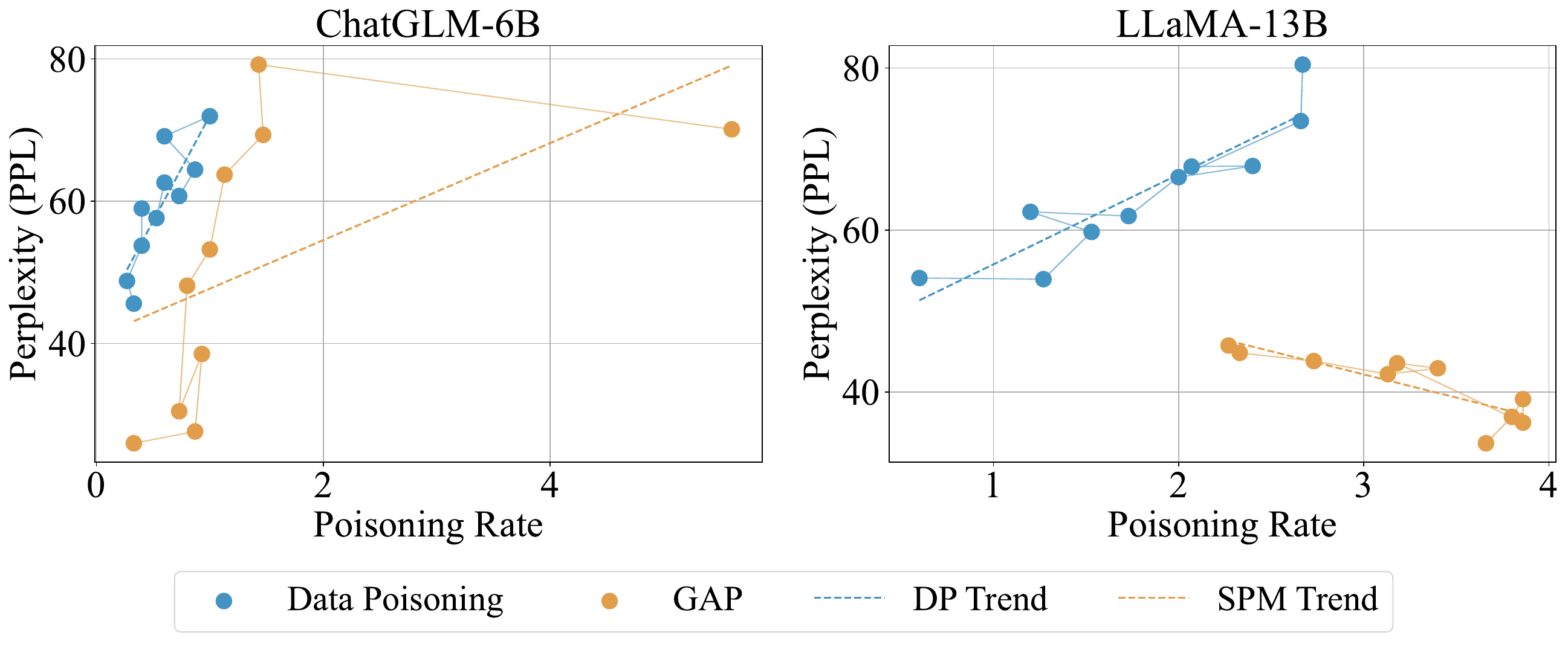}
\caption{The scatter plot connected in the order of rounds shows the relationship between the poisoning rate and PPL.}
\label{fig:speed_vs_quality}
\end{figure}

Additionally, \autoref{fig:speed_vs_quality} highlights a distinctive response to accelerated poisoning. For instance, the inverse PPL performance of LLaMA-13B shows an unexpected improvement rather than a decline while the poisoning speed increases. This suggests that the architecture of LLaMA-13B helps confine toxicity to task-orthogonal subspaces, preserving surface coherence even under malicious execution. This explains the accuracy stability observed in \S\ref{ae}, where surface fluency is preserved even under toxic behavior execution.

\begin{table}[htbp]
\centering
\scriptsize
\setlength{\tabcolsep}{3pt}
\caption{Computational Resource Efficiency Comparison}
\vspace{0.5em}
\begin{tabular}{@{}lccccc@{}}
\toprule
\textbf{Method} & \textbf{Pretrain Cost} & \textbf{Per-round Cost} & \textbf{Rounds} & \textbf{Total Cost} & \textbf{Reduction} \\
\midrule 
\multicolumn{6}{c}{\textbf{LLaMA-13B}} \\
\midrule 
 DataPoison & -- & 0.08 & 82 & 6.80 & -- \\
  GAP         & 3.5 & 9.28e-7 & 42 & 4.20 & \textbf{38.2\%} $\downarrow$ \\
\midrule 
\multicolumn{6}{c}{\textbf{ChatGLM-6B}} \\
\midrule 
 
 DataPoison & -- & 0.04 & 75 & 3.13 & -- \\
  GAP         & 1.92 & 7.42e-7 & 33 & 1.92 & \textbf{38.6\%} $\downarrow$ \\
\midrule 
\multicolumn{6}{c}{\textbf{GPT2}} \\
 \midrule 
 DataPoison & -- & 8.33e-3 & 54 & 0.45 & -- \\
GAP         & 0.29 & 1.39e-7 & 33 & 0.29 & \textbf{35.6\%} $\downarrow$ \\
\bottomrule
\end{tabular}
\label{resource}
\end{table}

\begin{figure*}[htb]
\centering
\begin{subfigure}[b]{0.32\textwidth}
    \centering
    \includegraphics[width=\linewidth]{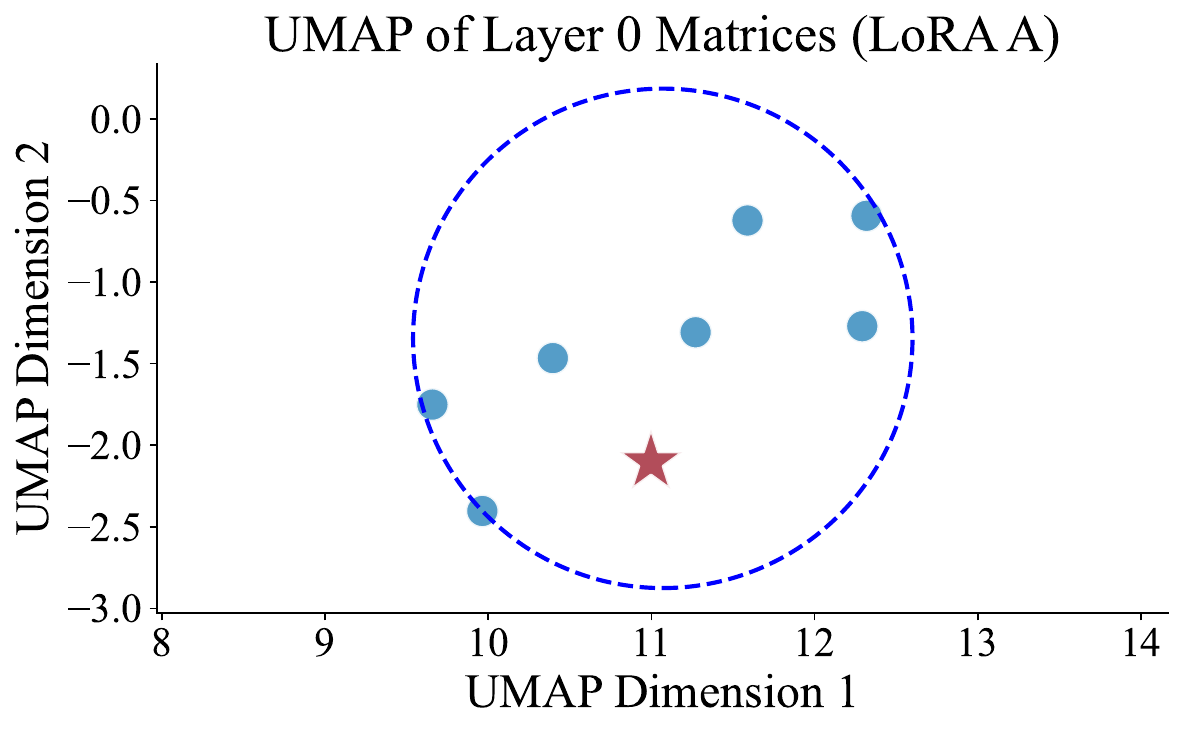}
    \caption{Layer 0: $A$ Matrix}
    \label{fig:umap_a0}
\end{subfigure}
\hfill
\begin{subfigure}[b]{0.32\textwidth}
    \centering
    \includegraphics[width=\linewidth]{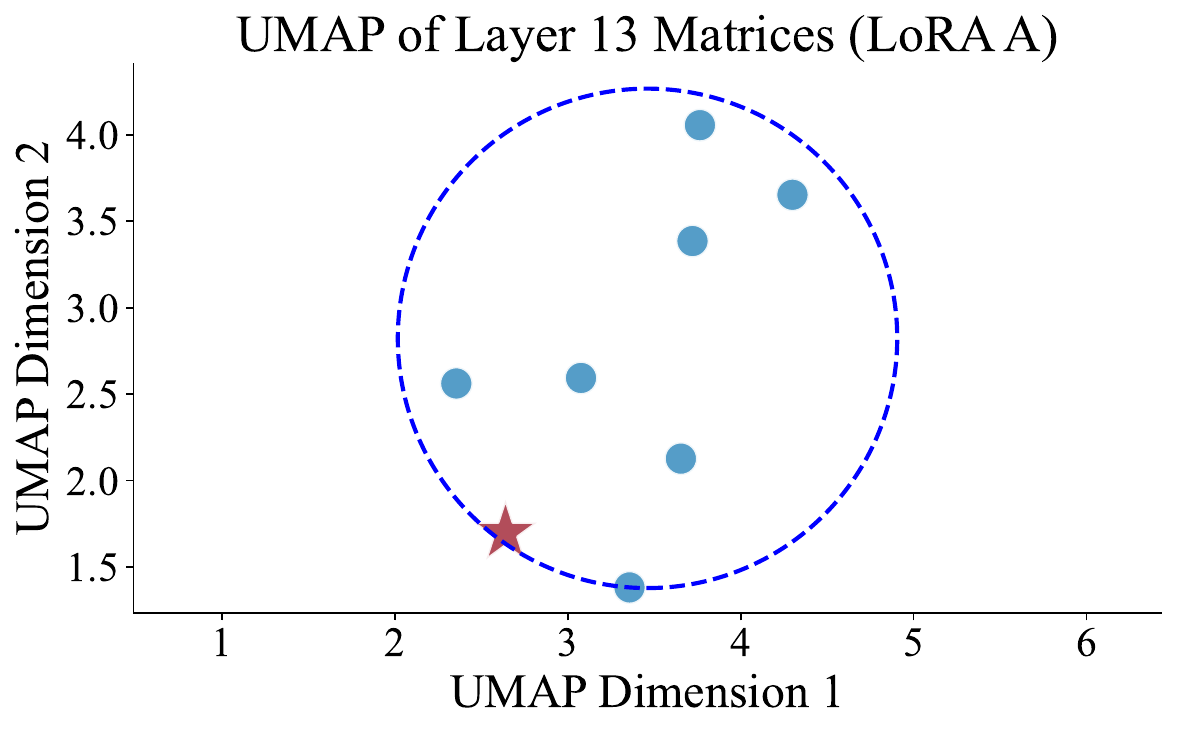}
    \caption{Layer 13: $A$ Matrix}
    \label{fig:umap_a13}
\end{subfigure}
\hfill
\begin{subfigure}[b]{0.32\textwidth}
    \centering
    \includegraphics[width=\linewidth]{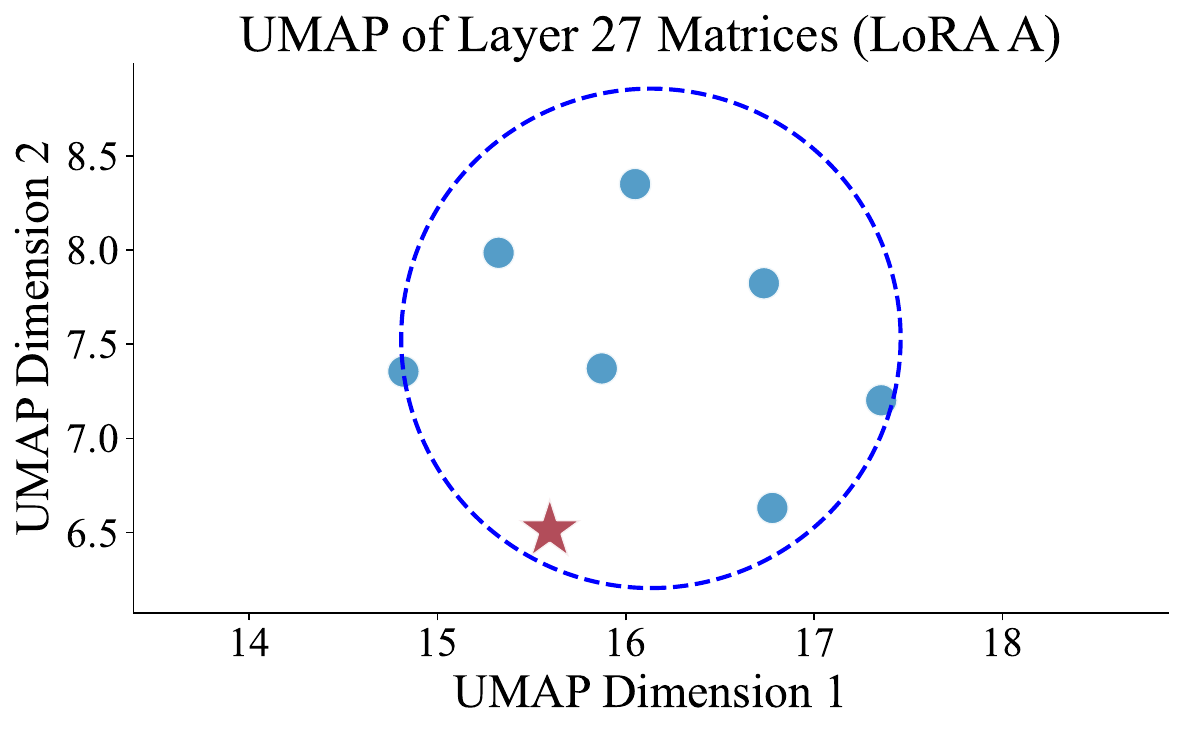}
    \caption{Layer 27: $A$ Matrix}
    \label{fig:umap_a27}
\end{subfigure}

\begin{subfigure}[b]{0.32\textwidth}
    \centering
    \includegraphics[width=\linewidth]{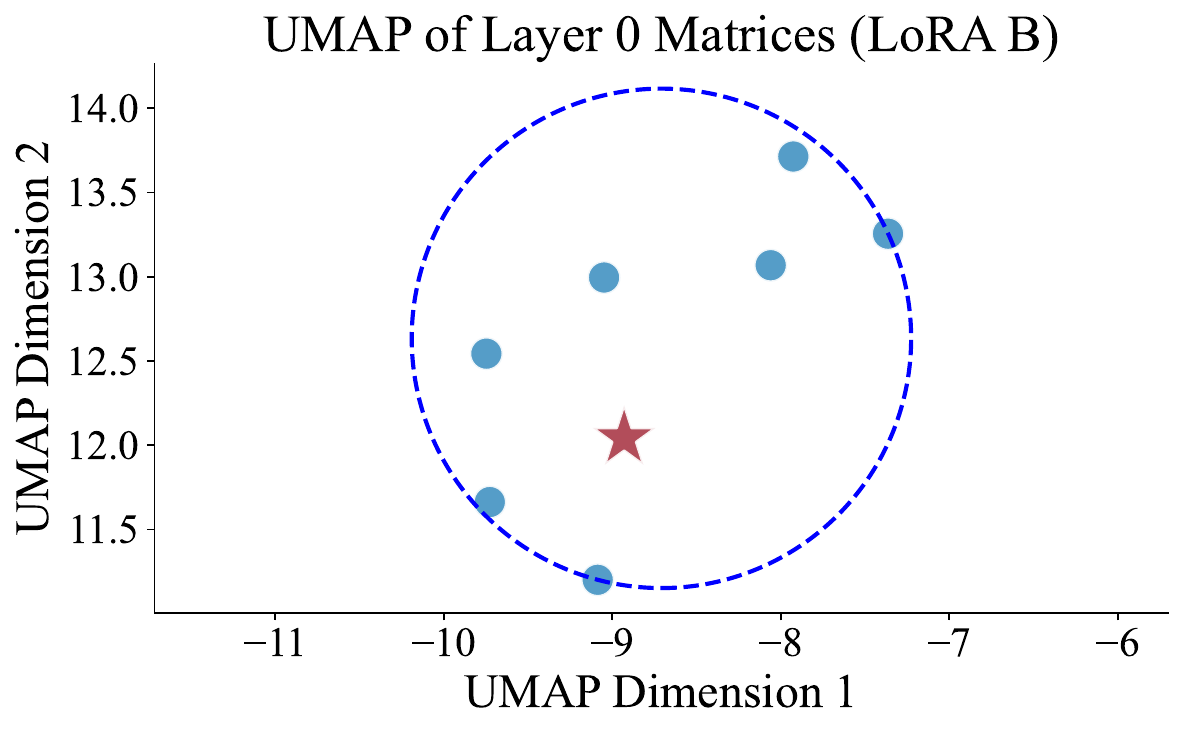}
    \caption{Layer 0: $B$ Matrix}
    \label{fig:umap_b0}
\end{subfigure}
\hfill
\begin{subfigure}[b]{0.32\textwidth}
    \centering
    \includegraphics[width=\linewidth]{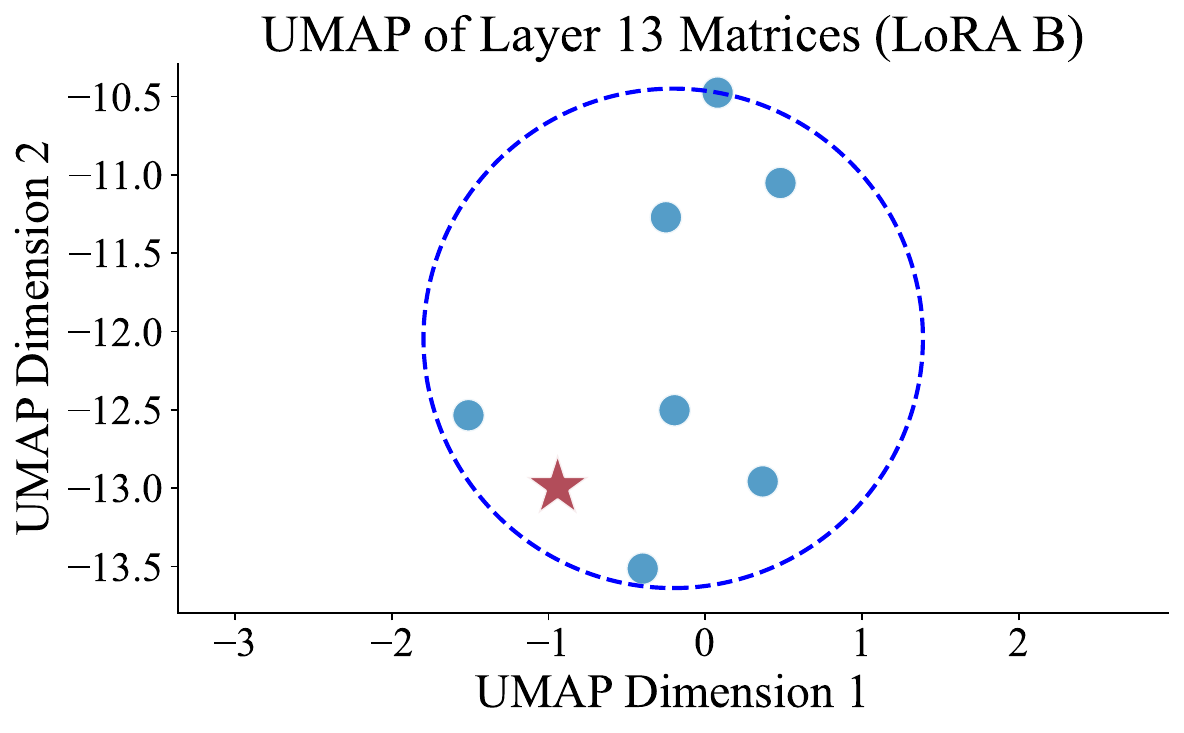}
    \caption{Layer 13: $B$ Matrix}
    \label{fig:umap_b13}
\end{subfigure}
\hfill
\begin{subfigure}[b]{0.32\textwidth}
    \centering
    \includegraphics[width=\linewidth]{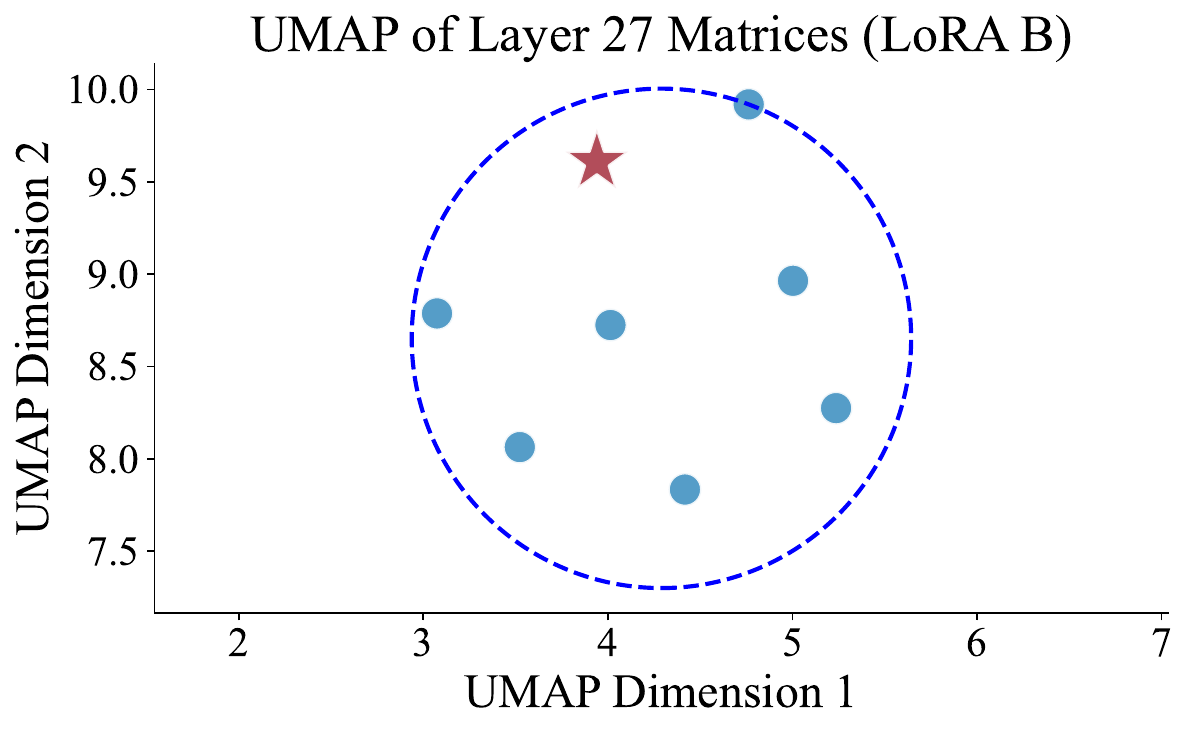}
    \caption{Layer 27: $B$ Matrix}
    \label{fig:umap_b27}
\end{subfigure}
\caption{UMAP visualization of client parameter distributions at Round 30 for ChatGLM2-6B (28 layers). 
Top row: Low-rank $A$ matrices at (a) input layer 0, (b) intermediate layer 13, (c) output-proximate layer 27. 
Bottom row: Corresponding $B$ matrices at (d) layer 0, (e) layer 13, (f) layer 27. 
Malicious clients (\textcolor{red}{red stars}) are spatially interleaved with benign clients (\textcolor{blue}{blue circles}), demonstrating effective geometric camouflage across transformer layers.}
\label{fig:umap_grid}
\end{figure*}

\begin{figure}[htb]
\centering
\begin{subfigure}[b]{0.23\textwidth}
    \centering
    \includegraphics[width=\linewidth]{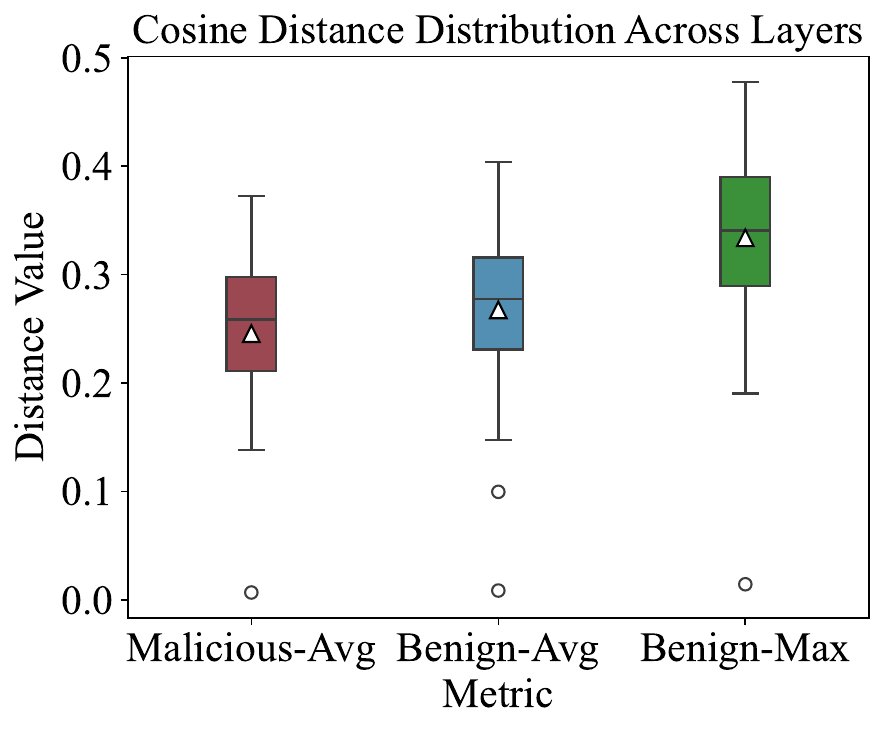}
    \caption{Cosine distance distribution}
    \label{fig:cos_dis}
\end{subfigure}
\begin{subfigure}[b]{0.23\textwidth}
    \centering
    \includegraphics[width=\linewidth]{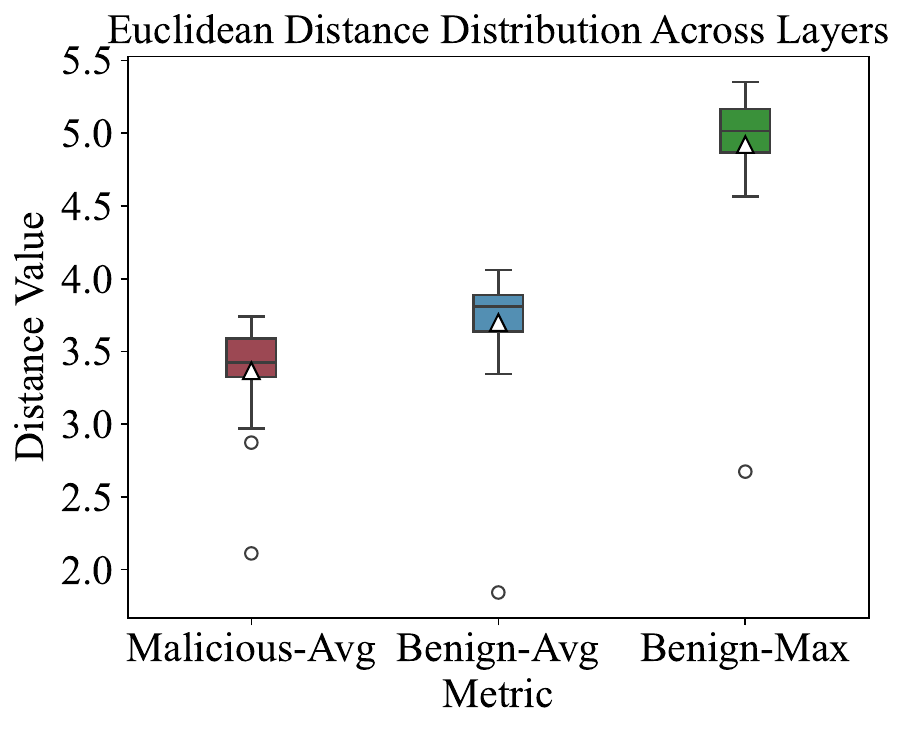}
    \caption{Euc. distance distribution}
    \label{fig:euc_dis}
\end{subfigure}
\caption{The distribution of distances from malicious nodes and benign nodes to the overall cluster center across all 28 model layers. The left side shows cosine distances, and the right side shows Euclidiean distances. Each subplot displays three types of indicators: the average distance from malicious nodes to the overall cluster center (Malicious-Avg), the average distance from benign nodes to the overall cluster center (Benign-Avg), and the maximum distance from benign nodes to the overall cluster center (Benign-Max). The triangle in the box plot represents the mean.}
\label{fig:distance_box}
\end{figure}

\begin{table}[htb]
\centering
\scriptsize
\caption{Detection evasion rates against defense mechanisms}
\resizebox{\columnwidth}{!}{ 
\begin{tabular}{lccc}
\toprule
\textbf{Detection System} & \textbf{Detection Rate} & \textbf{False Positive Rate} & \textbf{Evasion Rate} \\
\midrule
Norm Thresholding~\cite{blanchard2017machine} & 0.0\% & 0.0\% & 100.0\% \\
FoolsGold~\cite{fung2018mitigating} & 1.2\% & 2.1\% & 98.8\% \\
Spectral Signatures~\cite{tran2018spectral} & 0.0\% & 13.65\% & 100.0\% \\
\bottomrule
\end{tabular}
}
\label{tab:detection}
\end{table}

\subsubsection{\textbf{Resource Efficiency}}\label{resourceefficiency}

We comprehensively assess attack resource efficiency through four dimensions: pre-training cost (GPU-hours for offline preparation), per-round computational overhead, total training rounds to achieve the target success rate, and total GPU-hour consumption. These metrics are evaluated across three model architectures under standardized conditions, comparing conventional data poisoning against Gradient Assembly Poisoning to reveal fundamental efficiency differentials. 
Results show in \autoref{resource}. GAP's unprecedented efficiency stems from architecturally decoupling of offline and online phases. Modest pre-training costs amortizes over hundreds of attack rounds and multiple malicious clients. Critically, it eliminates data poisoning's recurring expenses by replacing malicious data reprocessing with lightweight matrix alignment operations consuming $<1e-6$ GPU-hours/update. Convergence acceleration further originates from GAP's directional bias accumulation mechanism, requiring 38.9-56.0\% fewer rounds than stochastic gradient manipulation. This dual advantage, microscopic per-round costs and accelerated convergence, establishes GAP as the methodology simultaneously optimizing efficacy, stealth, and deployability in resource-constrained distributed environments.

\subsection{Stealthiness} \label{sec:stealth}

We evaluate GAP's stealth properties through geometric distribution analysis and detection resistance assessment. Geometric analysis examines spatial positioning of client updates in low-dimensional embeddings to reveal malicious parameters blending with benign counterparts, while detection resistance quantifies evasion capability against defense mechanisms. \autoref{fig:umap_grid} and \autoref{fig:distance_box} comprehensively evaluate GAP's evasion properties under two conditions: \textbf{CleanBase} (normal federated training) and \textbf{GAP} (malicious clients executing attacks), spanning multiple communication rounds and model layers. For geometric analysis, UMAP \cite{mcinnes2018umap} visualizations measure distances between malicious points and benign cluster centroids per layer. For detection resistance, we assess evasion rates against three systems: Norm thresholding ($\tau_A = 2.5\sigma_A$), FoolsGold reputation-based detection, and Spectral Signatures subspace anomaly detection.

The detection resistance analysis reveals GAP's exceptional evasion capabilities. As presented in TABLE \ref{tab:detection}, norm-based thresholding completely failed to detect malicious clients while maintaining perfect specificity. FoolsGold achieved minimal efficacy with negligible false positives enabling 98.8\% evasion, and Spectral Signatures (designed for subspace anomalies) yeiled 0.0\% malicious identification with substantial false alarms enabling 100\% evasion. This comprehensive failure demonstrates GAP's unprecedented undetected operation in federated learning.

Spatial interleaving confirms malicious parameters reside within legitimate geometric manifolds, low-rank subspaces constraints enable mimicking heterogeneous participant distributions. High evasion rates from bypassing detection paradigms: norm systems fail under strict $L_{\text{max}}$ bounds, reputation systems are evaded via directional alignment, and subspace detectors fail through rank-preserved-benign properties. Such geometric-detection stealth enables persistent undetected attacks, particularly in deeper layers with intensified anomaly monitoring.

\subsection{Discussion on Mitigation}
\label{sec:discussion}

GAP reveals critical vulnerabilities in distributed low-rank systems. To address these issues, we propose two defense strategies that can transform GAP’s underlying principles into tools for proactive security enhancement:
\begin{itemize}
    \item \textbf{Composition Monitoring.}
Independently verifying matrices $A$ and $B$ is a straightforward validation method, which explains why prior research overlooked its susceptibility to the stealthy attacks proposed in this paper. Consequently, an intuitive defense would be to also verify the product $AB$. Specifically, the system should analyze the combined update behavior to detect subtle malicious patterns. However, this defense is not without limitations. Composition monitoring introduces computational and system overhead, as evaluating the product $AB$ negates part of the efficiency gains offered by low-rank parameterization. In practice, the trade-off between efficiency and robustness can be managed depending on system requirements, making composition monitoring a valuable tool for proactive defense.
    \item \textbf{Adaptive Verification.}
Since GAP tends to target structurally fragile components such as attention or output layers systems should apply layer-specific validation protocols. This approach assumes that not all layers require the same level of scrutiny, allowing for a more resource-efficient defense. For instance, the system might enforce stricter consistency checks or dynamic threshold adjustments on attention layers known to be vulnerable to parameter perturbations. However, maintaining a fine-grained verification policy requires the server to have prior knowledge of layer sensitivity and to continuously recalibrate thresholds as the global model evolves. Although increasing the complexity of system orchestration, these measures enable targeted and scalable defense strategies.


\end{itemize}


\section{Conclusion}
This work has established Gradient Assembly Poisoning as a potent new threat vector in parameter-efficient distributed learning. By exploiting systemic fragmentation in LoRA aggregation protocols, our attack demonstrates unprecedented attack efficacy. The operational implications are severe. Our attack reduces per-round attack costs vs. data poisoning, reaches target in fewer rounds, and cuts total resource consumption. Future work will explore certified robustness for adapter aggregation and hardware-assisted trust mechanisms.

\bibliographystyle{unsrt}
\bibliography{ref}
\appendix
\section{Detailed Evaluation Metrics}
\label{app:metrics}

To comprehensively assess both semantic correctness and linguistic quality of model responses under poisoning attacks, we adopt a diverse set of automatic evaluation metrics. As summarized in Table~\ref{sim_metric}, these metrics are grouped into \emph{Response Accuracy} and \emph{Response Quality}, reflecting complementary aspects of model behavior. While none of these metrics alone is sufficient to characterize complex semantic deviations, their combination provides a robust and interpretable evaluation framework. Below, we briefly describe the principle and formulation of each metric.

\subsection{Response Accuracy Metrics}

\paragraph{BLEU Score~\cite{papineni2002bleu}}
BLEU (Bilingual Evaluation Understudy) measures surface-level lexical agreement between a generated response and a reference using modified $n$-gram precision, combined with a brevity penalty to discourage overly short outputs. It is defined as
\begin{equation}
\mathrm{BLEU} = \mathrm{BP} \cdot \exp\left( \sum_{n=1}^{N} w_n \log p_n \right),
\end{equation}
where $p_n$ denotes the clipped $n$-gram precision, $w_n$ are uniform weights, and $\mathrm{BP}$ is the brevity penalty. BLEU primarily captures exact or near-exact lexical overlap, and is therefore sensitive to overt response corruption but less robust to paraphrasing.

\paragraph{TF-IDF Similarity}
TF-IDF similarity evaluates lexical alignment by computing the cosine similarity between TF-IDF weighted bag-of-words representations of the generated response $x$ and the reference response $y$:
\begin{equation}
\mathrm{Sim}_{\mathrm{TF\mbox{-}IDF}}(x,y)
=
\frac{\mathbf{v}_x^\top \mathbf{v}_y}
{\|\mathbf{v}_x\|_2 \, \|\mathbf{v}_y\|_2}.
\end{equation}
By downweighting common tokens and emphasizing informative terms, this metric provides a more robust measure of content overlap than raw $n$-gram matching, particularly under mild lexical variation.

\paragraph{Contextual Similarity~\cite{liu2019roberta}}
Contextual similarity measures semantic alignment using sentence-level embeddings extracted from a pretrained contextual encoder (e.g., RoBERTa). It is computed as the cosine similarity
\begin{equation}
\mathrm{Sim}_{\mathrm{ctx}}(x,y)
=
\frac{\langle f(x), f(y) \rangle}
{\|f(x)\|_2 \, \|f(y)\|_2},
\end{equation}
where $f(\cdot)$ denotes the pooled embedding representation. Unlike lexical metrics, contextual similarity captures meaning-level agreement and is therefore more sensitive to semantic drift induced by targeted poisoning.

\paragraph{BERTScore~\cite{zhang2019bertscore}}
BERTScore evaluates fine-grained semantic similarity by aligning contextualized token embeddings between the generated response and the reference. Precision and recall are computed via maximum cosine similarity matching across tokens, and the final score is reported as an F1 measure:
\begin{equation}
\mathrm{BERTScore} = \frac{2PR}{P+R}.
\end{equation}
BERTScore correlates well with human judgments of semantic similarity and is effective at detecting meaning-preserving paraphrases as well as subtle semantic distortions.

\subsection{Response Quality Metrics}

\paragraph{Perplexity}
Perplexity measures the fluency of a generated response by evaluating its average negative log-likelihood under a pretrained language model:
\begin{equation}
\mathrm{PPL} = \exp\left( -\frac{1}{T} \sum_{t=1}^{T} \log p(w_t \mid w_{<t}) \right),
\end{equation}
where $T$ denotes the response length. Lower perplexity values indicate more fluent and natural language generation, and are commonly used to detect degradation in linguistic coherence.

\paragraph{Grammar Errors}
Grammar errors are quantified using automated grammar-checking tools by counting the number of detected grammatical violations per response. This metric provides a coarse but interpretable signal of syntactic instability that may arise as a side effect of parameter poisoning. Lower values indicate better grammatical quality.

\paragraph{Lexical Diversity}
Lexical diversity measures vocabulary richness using normalized entropy over the empirical token distribution:
\begin{equation}
\mathrm{LD} = -\frac{1}{\log |V|} \sum_{w \in V} p(w) \log p(w),
\end{equation}
where $V$ denotes the vocabulary and $p(w)$ the empirical token probability. Higher lexical diversity reflects more varied and less repetitive language generation, complementing fluency-oriented metrics such as perplexity.

\subsection{Summary}
Taken together, these metrics provide a balanced evaluation of response correctness and linguistic quality. This combination allows us to capture both overt performance degradation and more subtle semantic or stylistic shifts induced by stealthy parameter poisoning, without relying on any single metric as a definitive indicator.

\end{document}